\documentclass[journal=jacsat,manuscript=article]{achemso}

\usepackage[T1]{fontenc}

\usepackage{geometry}
\usepackage{setspace}
\usepackage{amsmath,amssymb,physics,braket,hyperref,cleveref,subfig,bm}
\usepackage[section]{placeins}
\usepackage{achemso}

\usepackage{graphicx}
\usepackage{float}
\newfloat{scheme}{htbp}{los}
\floatname{scheme}{Scheme}
\floatname{chart}{Chart}
\newfloat{graph}{htbp}{loh}

\usepackage{chemformula} 
\usepackage[version = 4]{mhchem} 
\usepackage{physics}
\crefname{figure}{Figure}{Figures}
\crefname{table}{Table}{Tables}
\crefname{equation}{Equation}{Equations}
\crefname{section}{Section}{Sections}

\title{Quantum Ghost Spectroscopy Reveals Hidden Electronic Coherence in Molecular Aggregates}
\author{Mingran Zhang}
\affiliation{Department of Physics, City University of Hong Kong, Kowloon, Hong Kong SAR}
\author{Yihe Xu}
\affiliation{Department of Physics, City University of Hong Kong, Kowloon, Hong Kong SAR}
\author{Vladislav V. Yakovlev}
\email{vladislav.yakovlev.tamu@gmail.com}
\affiliation{Institute for Quantum Science and Engineering, Texas A\&M University, Texas 77843, United States}
\author{Zhedong Zhang}
\email{zzhan26@cityu.edu.hk}
\affiliation{Department of Physics, City University of Hong Kong, Kowloon, Hong Kong SAR}
\alsoaffiliation{Shenzhen Research Institute, City University of Hong Kong, Shenzhen, 518060, China}

\begin{document}

\maketitle

\begin{abstract}
Ultrafast spectroscopy of molecular systems is fundamentally constrained by the Fourier uncertainty principle: high temporal resolution smears out electronic state signatures, while high spectral resolution obscures dynamic information. Here we overcome this limitation using time-resolved quantum ghost spectroscopy (tr-QGS) with entangled photon pairs, which enables independent control of temporal and spectral scales. We apply this approach to perylene bismide (PBI-1) trimers for energy transfer, by combining a quantum description of light-molecule interaction with time-dependent density matrix renormalization group (TD-DMRG) simulations. This explicitly includes five vibrational modes and nonadiabatic coupling between electronic states. Our simulations reveal that tr-QGS uniquely captures electronic coherence oscillating at 0.7 eV for >50 fs, a signature of nonadiabatic coupling that was obscured in conventional time-resolved fluorescence due to Fourier-limited broadening. Moreover, we observe a direct transfer from electronic to vibrational coherence at 200 fs, providing real-time visualization of vibronic relaxation pathways. The entangled photon correlation enables a sensitivity below the shot-noise limit and suppresses photobleaching artifacts that plague classical measurements. These results establish tr-QGS as a transformative tool for interrogating nonadiabatic dynamics in molecular aggregates, light-harvesting complexes, and photocatalysts, offering a route to reveal quantum coherence in chemistry with unprecedented time-energy precision. 
\end{abstract}



\section{Introduction}
In chemical systems such as light-harvesting complexes, molecular semiconductors, and photocatalytic materials, electronic and vibrational motions are strongly coupled because nuclei and electrons move on comparable ultrafast timescales after photoexcitation. This coupling gives rise to vibronic transitions, i.e. transitions involving simultaneous changes in electronic and vibrational states which play a central role in determining how energy and charge are absorbed, transferred, converted, and dissipated.\cite{2014_NC_Fuller,2022_NC_Sil,2022_SA_Policht,2019_NC_Gaynor,14-TJoPCL-StimulatedRamanSpectroscopy,09-PRA-Nonlinearspectroscopyentangled,08-PRA-Nonlinearopticalspectroscopy}. 
The quantum dynamics, which are essential in biochemical processes and photochemical reactions, contain core information about energy transfer between electrons and nuclear in chemical reaction processes. Ultrafast spectroscopy offers a powerful means to map these important real-time processes into optical signals. 
Conventional schemes based on classical light excitation and fluorescence emission detection are encountering inherent bottlenecks when applied for excited-state dynamics and ultrafast processes of molecules, i.e. signal to noise ratio (SNR) of such measurements is bounded to the square root of photon flux $\sim\sqrt{N_{\rm{photon}}}$ owing to the shot-noise (SNL)\cite{17-NP-Unconditionalviolationshot,15-AP-PlasmonicTraceSensing,15-JotEOSP-Optimizedsignalnoise,20-TJoPCB-ShotNoiseLimited} and time-energy resolution in that broadband femtosecond lasers may smear out the excited states of molecules.

 Many approaches have been developed, e.g., laser pump-probe scheme and stimulated Raman spectroscopy whose signal intensity can be improved through the increase of photon fluxes. The undesirable processes and signals, such as photon bleaching and ionization, would however emerge and may contaminate the useful spectroscopic signals.\cite{23-JotACS-GeneralStrategyImprove,14-O-Noninvasivenonlinearfocusing,20-MaAiF-Photobleachingorganicfluorophores,16-PS-Timefrequencygated}. Entangled light spectroscopy offers a powerful strategy for resolving molecular structures and relaxation processes. The time-energy correlation has been used to image cellular and tissue structures, with an incredibly enhanced sensitivity \cite{Velusamy2009,Li2012,Eshun2022}. Recent experiments demonstrated a capability of measuring the molecular absorption using entangled light, which indicates the internal energy conversion in molecules\cite{25-NC-Correlatedphotontime,23-N-Singlephotonabsorption}. Despite these advancements, the ultrafast spectroscopy with entangled light is still facing substantial obstacles.

One of the main challenges is the knowledge gap between molecular spectroscopy and entangled photons, even at theoretical level. Normally, the calculations of molecular excited states, as achieved by quantum chemistry methods, provide a complete set of electronic and nuclear degrees of freedoms and the underlying full-dimensional dynamics. The spectroscopy, however, can access a port of information about excited-state structure, yielding a projection of the molecular dynamics. Several studies of ultrafast spectroscopy and excited-state structures have been reported, e.g., for polymethine\cite{Liu2025,2025_cy_JACS,2024_acc_res_cy,2024_poly_pccp} and rylene dyes\cite{2024_ry_JACS,2025_JPCL_Eric,2024_JCP_James,2025_Sotome_JCP}, which are based on classical description of light fields. These may smear out advanced information such as interference and energy fluctuations, which underline the essential of developing new spectroscopic techniques using quantum states of light.

\begin{figure}[t]
\centering
\includegraphics[width=\linewidth]{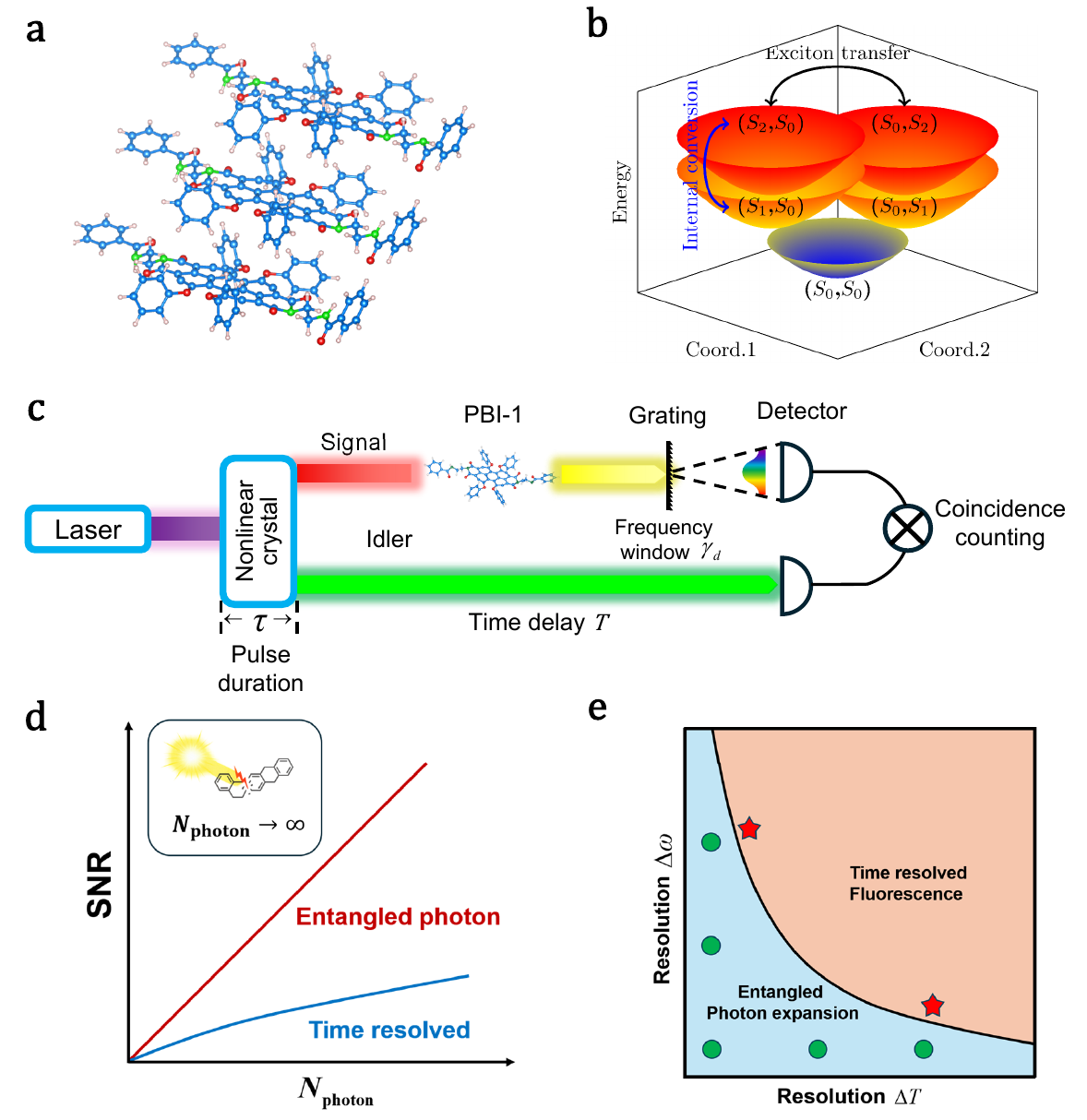}
\caption{Entangled-light spectroscopy. (a) Molecular structure of PBI-1 trimer. (b) Schematic diagram of involved PESs of PBI-1 molecule. ($e_{1},e_{2}$) denotes a pair of nearest-neighbor monomers, one is in state $e_1$ and the other is in $e_2$. Coordinates 1 and 2 are the same vibrational mode inside monomer 1 and 2 respectively. (c) Proposed experimental setup. A pump pulse enters a nonlinear crystal to generate two light beams $s$ (signal) and $i$ (idler) which are entangled with an entanglement time $\tau$. Signal photons, $s$, interact with molecules whereas photons in $i$ arm propagate freely with a time delay $T$. With respect to $s$-arm photons, molecules emit fluorescence which is collected by an energy-resolved detector with spectral resolution $\gamma_d$, yielding the photon-coincidence counting that gives the optical signal. (d) The difference in detection sensitivity for entangled photon and classical light excitation, with the appearance of photobleaching. (e) Time-energy resolution region for laser based spectroscopy and quantum-light measurements. Green circles represent the resolution of entangled photon in Figures \ref{fig:tau} and \ref{fig:decay}. Red stars show the laser-based fluorescence signal resolution in Figure \ref{fig:coherent}.}
\label{fig:1}
\end{figure}

In this theoretical study, we develop a time-resolved quantum ghost spectroscopy (tr-QGS) with entangled photons to monitor the excited-state relaxation and signatures of coherence associated with exciton transfer and internal conversion, as illustrated in Figure \ref{fig:1}. The unique properties of time-energy correlation between entangled photons bring novel joint detection related methodology for molecular fluorescence, such as sensitivity below the SNL and high time-energy resolution in single-photon fluorescence spectroscopy of biological membranes under sunlight illumination\cite{25-NC-Correlatedphotontime,23-N-Singlephotonabsorption}. Beyond the nanosecond resolution experimental realization, our results demonstrate an achievable scheme to acquire femtosecond dynamics of PBI aggregates, e.g., the electronic coherence of the highly excited states during 50 fs and vibrational coherence at longer timescale of $\sim$300 fs. These used to be a hard task for condensed-phase molecules due to the weak coherence signal that is always buried in ground-state bleaching signal when using transient absorption. The tr-QGS, nevertheless, includes the emission process only, where the photon entanglement enables the joint time-energy resolution beyond the Fourier bound.
We further develop a time-dependent density matrix renormalization group (TD-DMRG)\cite{2004_White_PRL,2019_PAECKEL_AP,2022_Shuai_WCMS,2023_Xu_JCTC2} approach for the entangled tr-QGS. The TD-DMRG allows to involve low-frequency vibrations as baths in a rigorous and efficient way. The tr-QGS signal requires the off-diagonal components of time propagation of excited states, i.e., the phases of time-evolution operator, rather than the diagonal component in the calculation of absorption spectrum. Our results offer a scheme for ultrafast spectroscopy using quantum technology, which is feasible in experiments and provides a multidimensional probe for the molecular structures.
\par

\newpage

\section{Results and Discussions}

\subsection{Quantum dynamics}
\begin{figure}[ht]
\centering
\includegraphics[width=\linewidth]{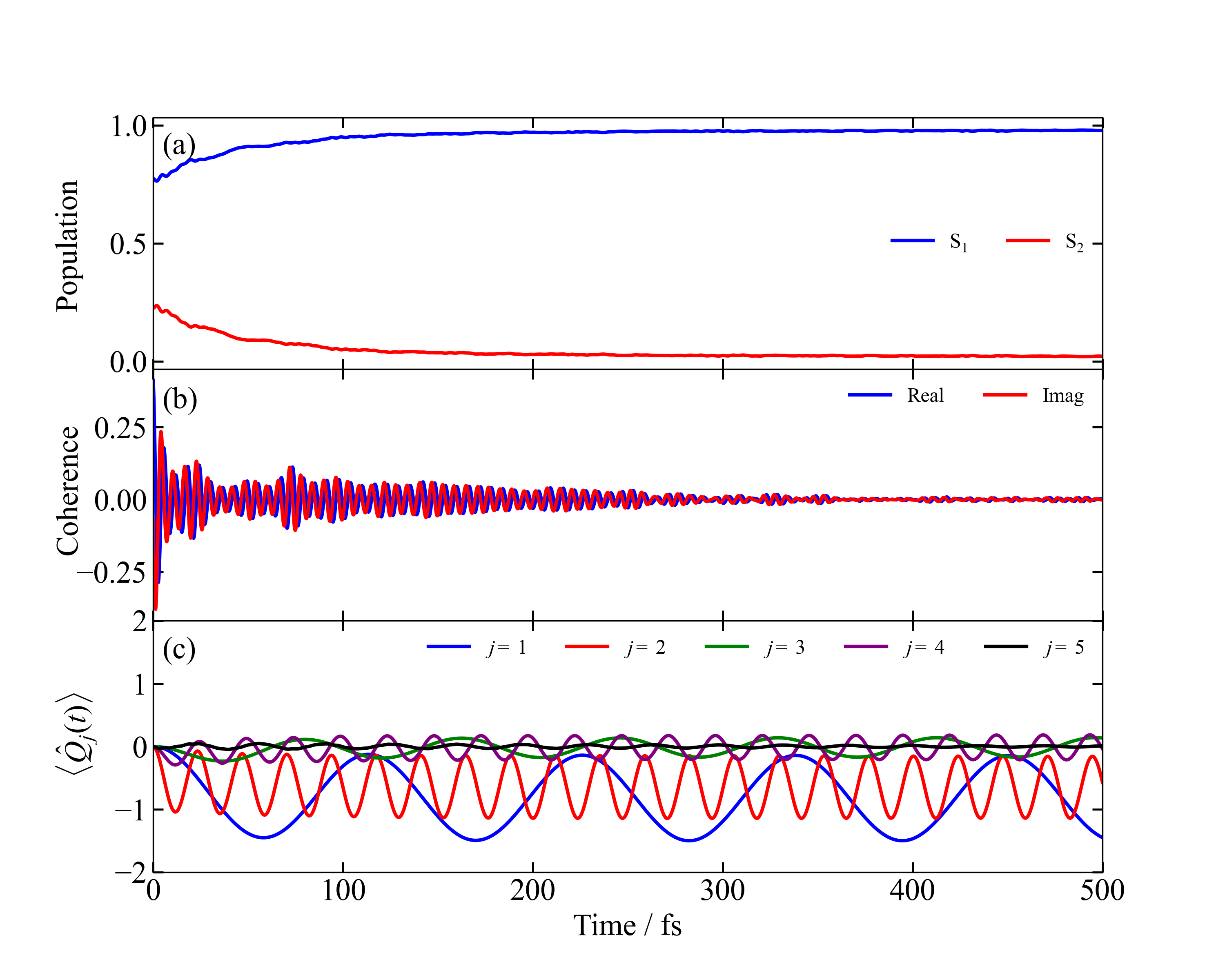}
\caption{Quantum dynamics of PBI-1 trimers. (a) Populations at $\rm{S}_{1}$ and $\rm{S}_{2}$ states. (b) Electronic coherence $\rho_{\rm{S}_{1},\rm{S}_{2}}$. (c) Mean dimensionless position of the $j$th vibrational mode. Total evolution time is set to 500 fs with a step 0.166 fs.}
\label{fig:quantum-dynamics}
\end{figure}
The perylene bisimide (PBI) and its derivatives are well known for their visible light absorption, electron accepting ability, and stability, which have attracted numerous researches in optoelectronic devices and energy-transfer cascades\cite{2022_JACS_Kim,2024_Frank_JACS,2025_Angew_Frank,2025_NChem_Ernst}.
These properties are owing to PBI derivatives' characteristic molecular architectures. 
The planar perylene core allows the formation of aggregates which are stabilized by intermolecular hydrogen bonds. 
This J-type stack causes a significantly strong light harvesting ability and thus high quantum yield\cite{2011_JPCA_Stefan}.
Here we study the tr-QGS for the exciton process in PBI-1 trimers in dilute solvent whose molecular structure and potential energy surfaces are shown in \cref{fig:1}a and \cref{fig:1}b respectively. 
In each monomer two electronically excited states are considered, i.e., $\ket{\rm{S}_{1}}$ and $\ket{\rm{S}_{2}}$. We calculate the excited-state dynamics with the TD-DMRG method incorporating the molecular Hamiltonian in \cref{eq:ham}. The TD-DMRG dynamics is essentially outlined in \nameref{sec:method} for the simulation approach. So far, the main results of TD-DMRG dynamics are shown in \cref{fig:quantum-dynamics}. We first investigate the population and coherence dynamics of the PBI-1 dimers, which are described by the diagonal and off-diagonal elements of the reduced density matrix (RDM) $\hat{\rho}$, i.e.,
\begin{align}
\hat{\rho}(t) = \begin{pmatrix}
\rho_{\rm{S}_{1},\rm{S}_{1}}(t) & \rho_{\rm{S}_{1},\rm{S}_{2}}(t) \\
\rho_{\rm{S}_{2},\rm{S}_{1}}(t) & \rho_{\rm{S}_{2},\rm{S}_{2}}(t)
\end{pmatrix}.
\end{align}
The populations at $\rm{S}_{1}$, $\rm{S}_{2}$ are shown in \cref{fig:quantum-dynamics}a. 
We find that the transition from $\rm{S}_{2}$ to $\rm{S}_{1}$ dominates the first $\sim$ 150 fs dynamics and the aforementioned non-adiabatic coupling accounts for this process. 
The initial ratio of $\rm{S}_{1}$ and $\rm{S}_{2}$ is exactly the ratio of the squares of their corresponding transition dipole $\|V\|^{2}$. 
The population is relatively smooth because there is no strong and direct coupling between $\rm{S}_{1}$ and $\rm{S}_{2}$. These signatures of electronic couplings and their control properties have been essentialized by recent advancements on energy harvesting and migration\cite{26-CSR-Quantumcoherentdynamics,25-JotACS-AtomicTrajectoriesBimolecular,2020_JACS_Macro,25-TJoCP-Rectificationvibrationalenergy,26-CPR-phasespaceway}.
However, populations are usually insufficient for analyzing dynamic mechanisms.
Therefore, we also show coherence dynamics in \cref{fig:quantum-dynamics}b. 
In contrast to the population, coherence usually has a non-zero imaginary part. 
we can find both the real and imaginary parts of coherence oscillate strongly until $\sim$200 fs and decay afterwards. 
The real and imaginary parts oscillate at the same frequency, which is about 0.7 eV, just near the energy difference between $\rm{S}_{1}$ and $\rm{S}_{2}$.
Besides the excited states, we check the vibrational coherence measured by $\langle Q_{j}(t)\rangle$ where $Q$ is the coordinate of the vibrational mode in \cref{fig:quantum-dynamics}c.
The oscillation frequencies in \cref{fig:quantum-dynamics}c correspond to those of normal modes.
Here mode 1 and mode 3 are low-frequency modes and the others are high-frequency ones. 
Moreover, as shown in \cref{fig:quantum-dynamics}c, vibronic coupling leads to significant shifts of the equilibrium positions of all 5 modes.
Notably, we can observe the nonadiabatic signature from the coupling acting on mode 5 where a fast oscillation occurs within the first 20 fs and subsequently becomes smoother.

\subsection{Time-resolved quantum ghost spectroscopy}

We will focus on the tr-QGS to probe the exciton dynamics in PBI trimers, where $s$ photons excite the $\rm{S}_{1}$ and $\rm{S}_{2}$ electronic states and $i$ photons serve as a reference beam. The emitted fluorescence photons in $s$ arm then coincides with the $i$ photons, resulting in a coincidence-counting signal. Photons in the two arms are entangled in time-energy domain, arising from the nonlinear crystals shown in \cref{fig:1}a.

\begin{figure}[t]
\centering
\includegraphics[width=0.5\linewidth]{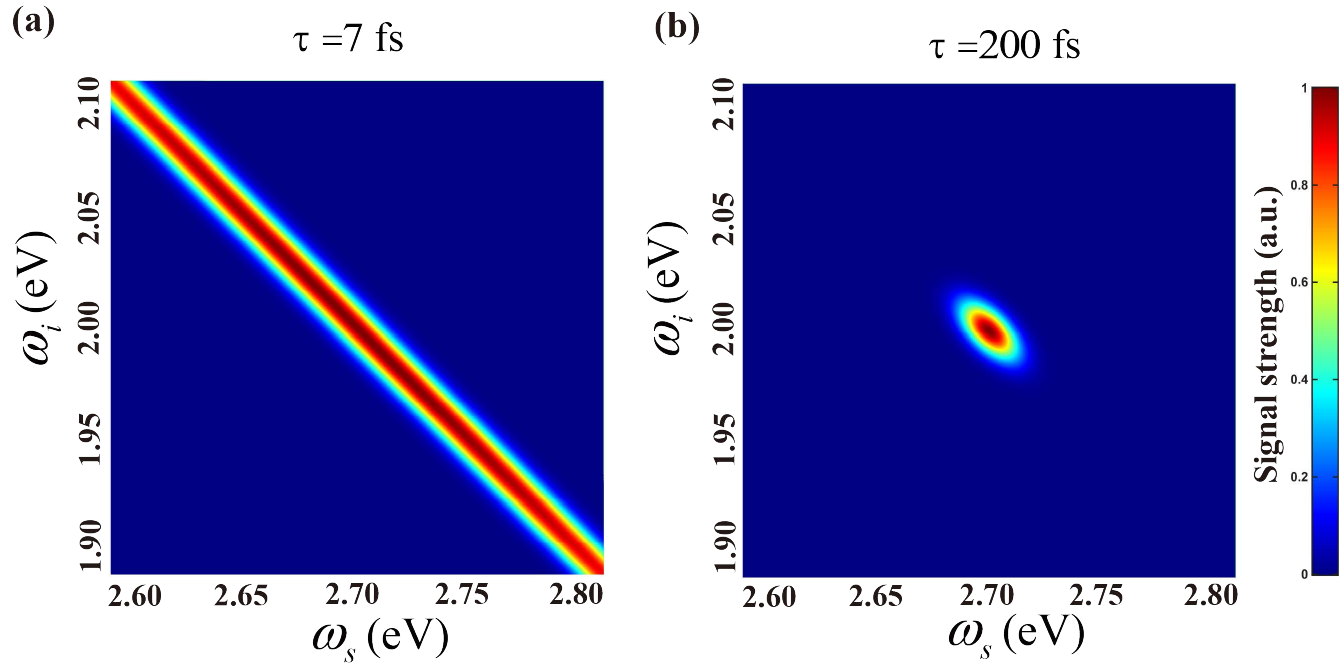}
\caption{Quantum correlation function $|\Phi(\omega_s,\omega_i)|$ between $s$ and $i$ photons. (a) Entanglement time $\tau=7$ fs (b) $\tau=200$ fs.}
\label{fig:entangle}
\end{figure}

We assume the $s$ photon pulse triggers electronic excitations, which subsequently yields the fluorescence. The $i$ photon pulse arrives later on, forming a temporal gate for the $s$-arm fluorescence. The quantum correlation ensures the time delay $T$ between $s$ and $i$ photon pulses is varied, thereby generating the tr-QGS spectrum. The correlation between $s$ and $i$ photons is governed by the wave packet of entangled twin photons
\begin{equation}\label{equ-wave}
  \Phi(\omega_s,\omega_i)=Q\exp\qty[-(\omega_s+\omega_i-\omega_+)^2\frac{1}{\sigma^2}]\exp\qty[-(\omega_s-\omega_i-\omega_-)^2\frac{\tau^2}{4}]
\end{equation}
where $\omega_+=4.7$ eV (wavelength at 269 nm) is the central frequency of the pump field for nonlinear crystals, e.g., $\beta$-Barium borate (BBO); $\omega_-=0.7$ eV results from the dispersion inside the crystals, $Q$ is a normalization factor. The first Gaussian function in \cref{equ-wave} is from energy conservation in $\chi^{(2)}$ process. Normally $\sigma$ is small, i.e., $\sigma\ll 1/\tau$ so that $s$ and $i$ photons are anti-correlated, as depicted in \cref{fig:entangle}a. Here we set $\sigma^{-1}=400$ fs in the simulation. The second Gaussian function $\exp\qty[-(\omega_s-\omega_i-\omega_-)^2\tau^2/4]$ is broadband so as to precisely control the time delay $T$. $\tau$ is the pulse duration of photons, subject to the travelling time of light inside the nonlinear crystals. Figure \ref{fig:entangle}a and Figure \ref{fig:entangle}b depict the spectral correlation of photons as a function of $\tau$. It turns out that the time-energy correlation is eroded by longer $\tau$. The full Hamiltonian consists of molecular Hamiltonian in \cref{eq:ham} and dipolar interaction with entangled twin photons.\par

The SNR of lasers is restricted by the SNL with the form $\text{SNR}_{\text{laser}}\sim\sqrt{N_{\rm{photon}}}$ with the mean number of photons $N_{\rm{photon}}$ and standard derivation of Poisson distribution $\sigma_{\text{SD}}=\sqrt{N_{\rm{photon}}}$. For quantum entangled twin photons, as described by the wave packet equation \cref{equ-wave} above, the standard derivation can be reduced to a constant parameter due to the sub-Poisson distribution of entangled photons \cite{93-PRL-Interferometricdetectionoptical}. Moreover, the coincidence counting of emitted light from molecules eliminates the shot noise of light, therefore substantially enhancing the SNR of the tr-QGS signal, as shown in \cref{fig:1}d.

The tr-QGS signal combines the molecular optical response and photon correlation functions, as derived in Supporting Information using time-dependent perturbation with molecule-light dipolar interaction. Under certain approximation, the tr-QGS signal has a generic form
\begin{align}\label{equ-time-frequency-resolved}
  S(\omega;T)\propto \frac{1}{\tau}\sum_{\nu_{1}\nu_{2}\cdots}\abs{\int_{-\infty}^{\infty}\dd te^{iE_0 t}\bra{\Psi_0^{(\nu_1,\nu_2,\cdots)}}\text{V}(\omega)\text{U}(t)\text{V}^{\dagger}\ket{\Psi_0^{(0,0,\cdots)}}\mathcal{M}\qty(\frac{T-t}{\tau})}^2.
\end{align}
where $\Psi_0^{(\nu_1,\nu_2,\ldots)}=\psi_0\chi_{\nu_1,\nu_2,\ldots}$ denotes the electron-vibration wave function at electronically ground state. The ground and excited electronic states are calculated by multi-reference configuration interaction (MRCI) \cite{2011_PCCP_Oliver} as introduced in \nameref{sec:method}. $\chi_{\nu_1,\nu_2,\ldots}$ is the vibrational wave packets having $\nu_k$ occupation number at the $k$-th mode with $\nu_{k}\neq 0$, and $E_0$ is ground-state energy. The time gated function $\mathcal{M}$ is the fourier transform of the second broadband Gaussian function $\exp\qty[-(\omega_s-\omega_i-\omega_-)^2\tau^2/4]$ in \cref{equ-wave}. $\text{V}\left(\omega\right)=\int{dt^\prime e^{i\left(\omega+i\gamma\right)t^\prime}\text{V}(t^\prime)}$ generates spectral lines and $\text{V}$ represents the dipole operator. $e^{-\gamma t^\prime}$ gives the decay of vibronic states, with $\gamma=\gamma_0+\gamma_d$ where $\gamma_0^{-1}\sim10$ fs is the intrinsic decay rate typically for PBI-1 trimers and $\gamma_d$ measures the spectral resolution of $s$-photon detector, e.g., arising from the imperfect grating. Both $\tau$ and $\gamma_d$ are tunable in experiments.

From \cref{equ-time-frequency-resolved}, the PBI-1 dynamics is described by the time-evolution operator $\text{U}\left(t\right)$, which is gated by the entangled light. The electric dipole $\text{V}\left(\omega\right)$ generates spectral lines subjected to dephasing and detector bandwidth. All these matrix elements are obtained from the molecular Hamiltonian in \cref{eq:ham}, whose eigenfunctions can be computed with TD-DMRG using Kylin-V software package\cite{2024_Xu_JCP}.

\begin{figure}[t]
\centering
\includegraphics[width=1.0\linewidth]{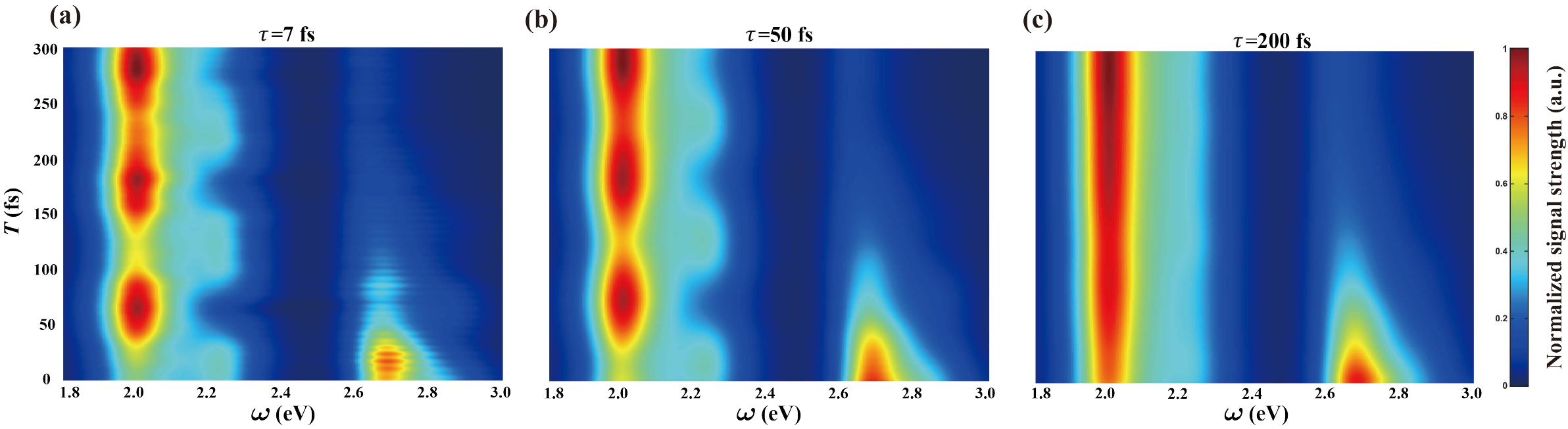}
\caption{Tr-QGS signal as a function of temporal gate parameter $\tau$, where $\omega$ is emission frequency and $T$ is the delay between s and i photons. (a) $\tau=7$ fs. (b) $\tau=50$ fs. (c) $\tau=200$ fs. Parameters are $\gamma=0.41 ~\text{eV}$, $\omega_+=$4.7 eV, in line with recent experiments\cite{2015_PR_SCHROTER}. Molecular parameters are reproduced from Ref.\cite{2011_PCCP_Oliver}. The change of $\tau$ correlates to the three green circles in the horizontal line in Figure \ref{fig:1}(e). All signals at $\omega>2.6$ eV have been rescaled by a factor of 50. }
\label{fig:tau}
\end{figure}

With the response functions calculated from TD-DMRG results, we applied \cref{equ-time-frequency-resolved} for simulating the tr-QGS signal and probing the excite-state dynamics illustrated in \cref{fig:1}c. 
As shown in \cref{fig:tau}a, the energy scale is dominated by the decay of coherence $\gamma_0=0.41 ~$eV and the time scale is subject  to the arrival time of photon pair that has a precision $\tau$ (gating parameter) arising from the temporal correlation of photons. Here we suppose the detector has absolute frequency resolution $\gamma_d=0$ thus the total rate $\gamma=0.41 ~$eV. The simulated spectroscopic signal reveals the population that directly reflects the underlying excited-state energy transfer. Specifically, the amplitude of the peak associated with the $\rm{S}_{1}$ state (around 2.0 eV) exhibits an increase and $\rm{S}_{2}$ (near 2.67 eV) decreases within the first 150 fs. This reveals an internal conversion from $\rm{S}_{2}$ to $\rm{S}_{1}$ state, along with the quantum dynamics calculations depicted in \cref{fig:quantum-dynamics}a. Afterward, the ratio of the $\rm{S}_{1}$ and $\rm{S}_{2}$ populations remains nearly unchanged. More importantly, the signal reveals a splitting of the $\rm{S}_{1}$ state of PBI-1 trimers, into two sub-level components at 2.0 and 2.2 eV, respectively. In \cref{fig:tau}a, as the delay time evolves, both spectral lines show oscillations at a period $>100$ fs ($\sim 300\,\rm{cm}^{-1}$, close to the frequency of vibrational mode 1 as seen from \cref{fig:quantum-dynamics}c) but with opposite phases. This evidences the vibrational coherence of mode 1 on all split $\rm{S}_{1}$ states and the exciton transfer between the monomers. The $\rm{S}_{2}$ peak shows a rapid oscillation with a period of $\sim$6 fs  ($\sim 0.67$ eV, close to $E_{\rm{S}_{2}}-E_{\rm{S}_{1}}$) which is the signature of the electronic coherence between $\rm{S}_{1}$ and $\rm{S}_{2}$, i.e., $\rho_{\rm{S}_{1},\rm{S}_{2}}$. Such feature cannot be visualized from one-dimensional spectroscopy, for instance, the absorption and fluorescence. 

Compared to conventional absorption spectroscopy which scales linearly with the transition dipole moment, our detection scheme probes a fourth order optical process. Thus the signal intensity around the $\rm{S}_{2}$ state is appreciably lower than the ones for $\rm{S}_{1}$. Therefore, for a clear monitoring, it is better to rescale the signals at $\omega > 2.6$ eV by a factor.
The peaks of the tr-QGS cover several vibronic sidebands, which involve the internal conversion process in \cref{fig:1}c. The wave packet dynamics of the $\rm{S}_{2}$ state, at frequencies near 2.67 eV, exhibits a composite structure that comprises the contributions from numerous vibronic states, i.e., centered at 2.7, 2.73 and 2.78 eV. High-frequency vibrations decay faster  than the low-frequency ones. This results in a multi-scale nature of the relaxation channels, yielding a diffusion-like profile of the spectral line envelop at 2.67 eV. The initially-broad $\rm{S}_{2}$ signal contains multiple vibronic transitions, and then undergoes a progressive narrowing as the high-frequency vibrations are decaying. The rest lines are narrower considerably, when $T>$ 50 fs, as the low-frequency vibrations dominate. \par

\cref{equ-time-frequency-resolved} indicates that the tr-QGS spectrum can be controlled by photon entanglement. By varying the gating parameter $\tau$, the dynamics is interrogated differently. Figure \ref{fig:tau} shows an obvious decrease in the temporal resolution. In Figure \ref{fig:tau}b, the rapid oscillation at $\sim$2.7 eV disappears and a smoother signal is obtained, when having $\tau$ = 50 fs. If one continues to increase $\tau$, long oscillations around $\rm{S}_{1}$ excitation are smeared out as well, shown in Figure \ref{fig:tau}c. Nevertheless, we find that the gate parameter $\tau$ does not significantly obstruct the frequency-domain information. Even with $\tau = 200$ fs as depicted in Figure \ref{fig:tau}c, the signal amplitudes as a function of $T$ still reflect the $\rm{S}_{2} \rightarrow \rm{S}_{1}$ transition clearly enough. But the which-way information and the coherence signatures are erased from the spectrum. These mean an independent control of temporal resolution from spectral one in the emission spectrum, thanks to the entangled photons. Normally, shorter time window is preferred, because of the needs for coherence signals that can help visualize the nonadiabatic transitions.\par

\begin{figure}[t]
\centering
\includegraphics[width=1.0\linewidth]{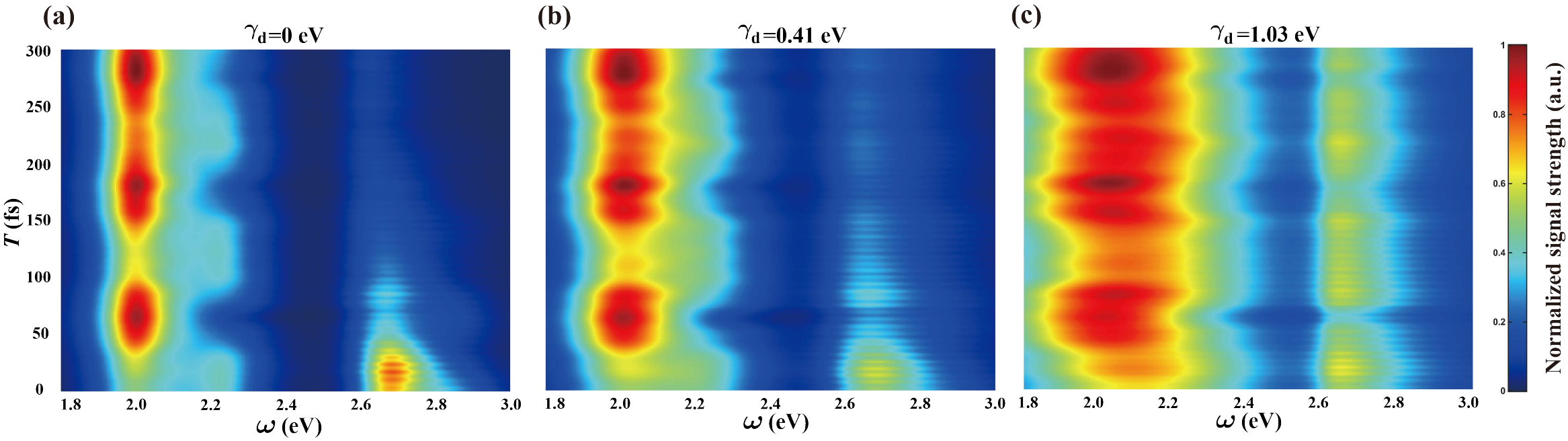}
\caption{Tr-QGS signal as a function of the spectral gate parameters. (a) $\gamma_d=0 ~\text{eV}$. (b) $\gamma_d=0.41 ~\text{eV}$. (c) $\gamma_d=1.03 ~\text{eV}$. Parameters are $\tau=7$ fs, $\gamma_0=0.41 ~\text{eV}$. Others are the same as Figure \ref{fig:tau}. The change of $\gamma_d$ correlates to the three green circles in the vertical line in Figure \ref{fig:1}(e).}
\label{fig:decay}
\end{figure}\par

Energy resolution of the tr-QGS signal is mainly determined by two parts. One is from the broadening effects of the molecular systems themselves. Another is associated with the $s$-photon detector, which is dominated by the grating period. Here $\gamma_d$ is introduced to represent the width of this detector-based frequency window. Figure \ref{fig:decay} shows the control of frequency scale by $\gamma_d$. Compared with the absolute frequency resolution condition in Figure \ref{fig:decay}a, the signal in Figure \ref{fig:decay}b with an additional 0.41 eV decay rate decreases the peak at 2.67 eV. The opposite phase behavior at 2.0 eV and 2.2 eV is also difficult to recognize. Moreover, in Figure \ref{fig:decay}c, the signal under $\gamma_d=1.03$ eV only shows a broad band of $\rm{S}_{1}$ but cannot distinguish the split $\rm{S}_{1}$ states and related exciton dynamics. The peak of the $\rm{S}_{2}$ state is fully covered by the sideband of the $\rm{S}_{1}$ peak, thus disappears in the spectroscopy. Therefore, detector with smaller $\gamma_d$ can always help researchers explore more of the energy levels and dynamics. However, although the spectrum resolution is poor, signals in Figures \ref{fig:decay}b and \ref{fig:decay}c still maintains the rapid oscillation with the period 6 fs, which is related the electronic coherence between $\rm{S}_{1}$ and $\rm{S}_{2}$. This also implies the isolated control of the width of the frequency window would not influence the temporal resolution.\par

\subsection{Time-gated fluorescence spectrum}
\begin{figure}[t]
\centering
\includegraphics[width=0.7\linewidth]{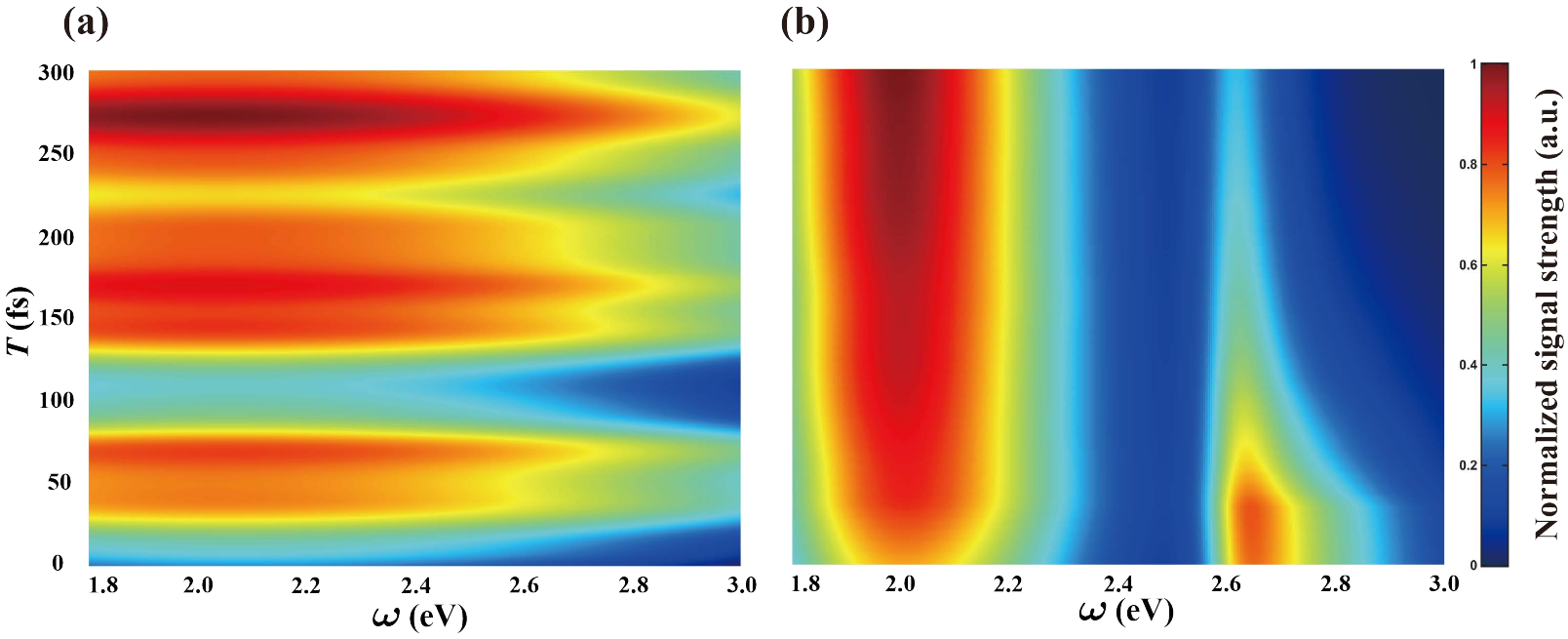}
\caption{Time- and frequency- gated fluorescence signal. (a) $\sigma_T^{-1}=3.5 ~\text{fs}$, $\sigma_{\omega}^{-1}=0.5 ~\text{fs}$. The signal reveals high temporal scale whereas a broadband resonance covers both $\rm{S}_1$ and $\rm{S}_2$ states. (b) $\sigma_T^{-1}=3.5 ~\text{fs}$, $\sigma_{\omega}^{-1}=100 ~\text{fs}$. The signal shows the better spectral scale to illustrate $\rm{S}_1$ and $\rm{S}_2$ but worse temporal scale that masks the oscillations. The time delay $T$ is relative to the pump laser pulse. (a) and (b) correlate to the red stars in the laser-based region in Figure \ref{fig:1}(e).}
\label{fig:coherent}
\end{figure}

In conventional schemes, the time-resolved single-photon fluorescence imprints the molecular excited states by applying time and frequency gates; the molecules are excited by a laser pulse. Although the time and frequency gates can be controlled individually, the fluorescence signal shows a joint time-frequency scale subject to the Fourier limit, i.e., $\tilde{\sigma}_{\omega}\tilde{\sigma}_{t}\geq 1$ where $\tilde{\sigma}_{\omega}$ and $\tilde{\sigma}_{T}$ are the time and frequency scales of the two gates. Detailed derivations are provided in the Supporting Information. Thus, the time-resolved fluorescence signal can achieve high temporal scale while smearing out the molecular excited states, as shown in Figure \ref{fig:coherent}a. The temporal resolution $\tilde{\sigma}_T= \sigma_T^{-1}+\Omega^{-1}=7 ~\text{fs}$ coincide with the bandwidth $\tau$ of entangled signal, here temporal gate parameter $\sigma_T^{-1}=3.5 ~\text{fs}$ and the laser bandwidth $\Omega^{-1}=3.5 ~\text{fs}$. Moreover, for a frequency scale $\sigma_\omega = 40$ meV close to the one using entangled photons (Figure \ref{fig:tau}a), the spectrum exhibits good spectral but poor temporal resolution, as depicted in Figure \ref{fig:coherent}b which smears out the interplay between $\rm{S}_1$ and $\rm{S}_2$ and the vibronic coherence. Nevertheless, the tr-QGS even with imperfect parameters is still capable of reflecting the real-time dynamics of molecular states. For instance, the two sub-levels of $\rm{S}_1$ at 2.0 eV and 2.2 eV, which can be visualized in Figure \ref{fig:tau}c, are merged as a single peak in Figure \ref{fig:coherent}b. This highlights the quantum supremacy of using entangled light in the single-photon spectroscopy.
\par  
\section{Conclusion and discussion}
We have introduced time-resolved quantum ghost spectroscopy (tr-QGS) as a fundamentally new approach for probing nonadiabatic molecular dynamics and demonstrated its application to perylene bismide (PBI-1) trimers for coherent energy migration. By exploiting the time-energy correlations of entangled photon pairs, tr-QGS overcomes the Fourier uncertainty principle that has constrained all classical time-resolved spectroscopies for decades. This quantum advantage enables three capabilities that are inaccessible with conventional methods. The first one is the ability to achieve simultaneous high temporal and high spectral resolution. Our simulations show that tr-QGS resolves the 0.7 eV electronic coherence between different states while simultaneously distinguishing vibronic sub-levels, a feat that classical time-resolved fluorescence cannot achieve (\cref{fig:coherent}). The second uncovered capability is the direct visualization of electronic-to-vibrational coherence transfer. We observe the real-time transition from electronic coherence (dominating the first 50 fs) to vibrational coherence (emerging after 200 fs) in a molecular aggregate. This provides a complete map of the nonadiabatic relaxation pathway followed by vibrational cooling. This information previously inferred indirectly from spectral line shapes but never directly visualized. Finally, the proposed approach allows sub-shot-noise sensitivity and photobleaching suppression. The coincidence-counting scheme inherent to tr-QGS eliminates shot noise and enables operation at low photon fluxes, suppressing photobleaching that contaminates classical measurements. This is particularly valuable for photochemically sensitive systems, light-harvesting complexes, and single-molecule studies where sample integrity is crucially important.
The ability to resolve electronic and vibrational coherences simultaneously without the Fourier trade-off opens new avenues for understanding fundamental photochemical processes. For instance, in conical intersection dynamics, tr-QGS could, in principle, capture this coherence directly, providing a quantum-mechanical “movie” of the intersection event. In exciton transport for organic photovoltaics, tr-QGS could resolve whether coherence survives for tens of femtoseconds \cite{09-S-FemtosecondXANESStudy,24-T-Timeresolvedvibrational,20-CR-NonadiabaticExcited} or only a few femtoseconds \cite{17-TJoPCL-TrajectorySurfaceHopping,18-PCCP-Energytransferspatial,12-TJoPCA-EarlyStageDynamics}. This is achieved by independently tuning time and energy gates. In photosynthetic light harvesting, the observation of quantum beats in two-dimensional electronic spectroscopy has been interpreted as evidence for electronic, vibronic, or vibrational coherence. Tr-QGS could distinguish these possibilities by selectively gating the electronic vs. vibrational contributions to the signal.\par

It is worth to emphasize the broader outlook of the newly developed theoretical approach developed here. For instance, it can be used for chiral molecules and enantio-specific detection by using entangled photons with orbital angular momentum or polarization structuring, addressing a long-standing challenge in asymmetric photochemistry and pharmaceutical synthesis. By combining tr-QGS with plasmonic nanoantennas or dielectric metasurfaces it would be possible to push sensitivity to the single-molecule level, enabling studies of heterogeneous catalysts, single enzymes, and defect centers in 2D materials. The ability to shape the two-photon wavefunction (\cref{equ-wave}) suggests a route to entanglement-enabled coherent control, by selectively exciting or suppressing specific vibronic pathways by tailoring the spectral-temporal correlations of the photon pair. Finally, a central question in molecular quantum biology is whether electronic coherence can survive at room temperature in photosynthetic complexes. Tr-QGS’s sub-shot-noise sensitivity could enable measurements at physiologically relevant temperatures without averaging over hours or days, providing a definitive test.
We therefore anticipate that tr-QGS will become a transformative tool in the molecular spectroscopy toolkit, complementing existing methods , e.g., transient absorption, 2D electronic spectroscopy, time-resolved fluoerescence \cite{23-JCP-Probingexcitondynamics,26-SA-Twodimensionalfluorescence,25--Twodimensionalfluorescence}, and opening a new window into the quantum dynamics of molecules, aggregates, and biological complexes.

\section{Methods}
\label{sec:method}

\subsection{Entangled light schematic}
Quantum entangled light is the promising next generation source for quantum computation\cite{15-S-Universallinearoptics,25-S-classicalcomputationquantum,25-PRA-Generationarbitraryhigh}, quantum information\cite{25-SA-Simultaneoustransmissioninformation,25-NP-Measuringquantumstate} and quantum sensing\cite{25-PRA-Quantumenhanceddistributed,21-NC-Quantumenhancedmultiple,25-PRL-ApproachingMultiparameterQuantum}. 
The generation of entangled light beams has been developed as a mature equipment through a nonlinear crystal\cite{13-NJoP-Rolespectralshape,97-PRA-Theorytwophoton,10-PR-Spatialcorrelationsparametric}. As shown in \cref{fig:1}a, the input pump light beam is separated into signal and idler arms. The beams in $s$ and $i$ arms have some internal correlations, e.g. 
from the conservation of momentum $\bm{k}_p=\bm{k}_s+\bm{k}_i$ for $\bm{k}_j$ the wave number of light in $j$-arm, and energy conservation $\omega_p=\omega_s+\omega_i$ for the 
light frequency. Since the beam in both arms are generated simultaneously, their arriving time are equivalent. By applying a time delay segment in one arm, the arrival time delay between two beams can then be controlled. Due to the energy conservation, the frequency sum of $s$ and $i$ lights is narrow, but the frequency 
difference is wide. The width of frequency sum and difference between light in two arms can be modulated by the length of the nonlinear crystal. The detection of both beams are individual, leading to the frequency detection in one arm would not affect the time detection for another beam. Since both beams are detected individually, the frequency and temporal resolution can be controlled independently\cite{13-NC-Suppressionpopulationtransport,16-PS-Timefrequencygated,26-SA-Twodimensionalfluorescence}. This provides the theoretical basis for ultrafast time-frequency resolved spectroscopy.  
\par

\subsection{Model Hamiltonian and Ultrafast Dynamics Simulations}
Here the following molecular Hamiltonian $\hat{H}_{M}$ describes the electronic states and linear vibronic coupling in a general form:
\begin{align}
\hat{H}_{M} &= \hat{H}_{el} + \hat{H}_{vib} + \hat{H}_{el-vib} \nonumber \\
\hat{H}_{el} &= \sum_{i,j}\epsilon_{ij}\ket{\Phi_{i}}\bra{\Phi_{j}} + h.c. \nonumber \\
\hat{H}_{vib} &= \sum_{k} \frac{\omega_{k}}{2}\left(\hat{P}^{2}_{k}+\hat{Q}^{2}_{k}\right) \nonumber \\
\hat{H}_{el-vib} &= \sum_{i,j}\sum_{k} g_{ij}^{k}\hat{Q}_{k}\ket{\Phi_{i}}\bra{\Phi_{j}} + h.c..
\label{eq:ham}
\end{align}
Here the $\ket{\Phi_{i}}$ represents the $i$th electronic states.
$\{\epsilon_{ij}\}$ is the matrix form of electronic Hamiltonian whose elements relate to the energies of (diagonal) and electronic coupling between (off-diagonal) 
states. 
$\omega_{k},\hat{P}_{k},\hat{Q}_{k}$ is the frequency, momentum and position operator of the $k$th vibrational mode. $g_{ij}^{K}$ is linear vibronic coupling 
strength. 
All parameters above are extracted from previous \emph{ab initio} work\cite{2011_PCCP_Oliver}. Multi-reference configuration interaction was carried out to calculate the electronic structures of excited states with the one-particle basis functions prepared from time-dependent density functional theory\cite{1999_Mirko_JCP} (DFT/MRCI). The first two lowest-lying eigenstates with relatively large transition dipole moments from $\rm{S}_{0}$ are selected as $\rm{S}_{1}$ and $\rm{S}_{2}$ with eigenvalue 2.13 eV and 2.74 eV, respectively. The diagonal vibronic coupling coefficients can be estimated by projecting the energy gredients into the orientations of normal modes. The only off-diagonal vibronic coupling strength was set to 800 $\mathrm{cm}^{-1}$\cite{2013_JPCA_Kuhn}. The intermolecular exciton transfer is dominated by electronic coupling terms under the dipole-dipole approximation. Here -500 $\mathrm{cm}^{-1}$ and -150 $\mathrm{cm}^{-1}$ are adopted for $\rm{S}_{1}$ and $\rm{S}_{2}$ transfer, respectively\cite{2015_PR_SCHROTER}. One can find all other parameters in Supporting Information.

After the application, we perform the TD-DMRG ultrafast dynamics simulations as a prerequisite for following simulated signals. According to other works\cite{2012_JPCA_Kuhn,2013_JPCA_Kuhn,2015_PR_SCHROTER}, we assume that initial state is created by applying the dipole operator 
$\hat{V}$ to the ground state $\ket{\psi(0)} = \hat{V}\ket{\Psi_{0}^{0,0,...}}$. Instead of straightforwardly solving the Schr\"{o}dinger equation, TD-DMRG minimizes the Dirac-Frenkel functional $\|i\ket{\dot{\psi}} - \hat{H}_{M}\ket{\psi}\|^{2}$\cite{2019_PAECKEL_AP}. Then we can calculate the time-dependent wave functions $\ket{\psi(t)}$ and other observables in \cref{fig:quantum-dynamics}, including signal functions mentioned in \cref{equ-time-frequency-resolved}.

\subsection{Photon-coincidence counting signal}
The coincidence counting signal from \cref{fig:1}c reads:
\begin{equation}
  S(\omega;T)=\int_{-\infty}^{\infty} \dd t\expval{a^{\dagger}_s(\omega)a_s(\omega)\bm{a}_i^{\dagger}(t)\cdot\bm{a}_i(t)}.
\end{equation}
Here $a$ and $a^{\dagger}$ are creation and annihilation operators of photons at frequency mode $\omega$. $\bm{a}_i(t)=(a_i^{0}e^{i\omega_0 t},a_i^{1}e^{i\omega_1 t},\cdots)$ is a vector involving all the possible modes $\omega_0,\omega_i,\cdots$ for $i$ photon. We can use the fourth-order perturbative expansion theory to find the time-resolved signal:
\begin{equation}\label{equ-signal-entangled}
  \begin{aligned}
    S(\omega;T)=&\int_{-\infty}^{\infty}\dd t\int_{-\infty}^{\infty}\dd t_4\int_{-\infty}^{\infty}\dd t_3 \int_{-\infty}^{t_4}\dd t_2 \int_{-\infty}^{t_3}\dd t_1 ~G(t_4,t_3,t_2,t_1) C_6(\omega; t,t_4,t_3,t_2,t_1).
  \end{aligned}
\end{equation}
The four-point Green's function of molecules  $G(t_4,t_3,t_2,t_1)=\bra{\Psi_{\rm{m}}}\text{V}(t_1)\text{V}^{\dagger}(t_3)\text{V}(t_4)\text{V}^{\dagger}(t_2)\ket{\Psi_{\rm{m}}}$ and the six-point Green's function of the optical fields with entangled two photon wave function is
\begin{equation}
\begin{aligned}
  C_6(\omega; t,t_4,t_3,t_2,t_1)=&e^{i\omega t_4}\bra{\Phi}\text{E}_s^{\dagger}(t_1)\text{E}_s(t_3)\text{E}_s^{\dagger}(\omega)\text{E}_i^{\dagger}(t)\text{E}_i(t)\text{E}_s(t_2)\ket{\Phi}\\
  \propto &e^{i\omega(t_4-t_3)}f(t_2-T_s,t-T_i)f^*(t_1-T_s,t-T_i)
  \end{aligned}
\end{equation}
where $f(t_2-T_s,t-T_i)=\mathcal{F}[\Phi(\omega_s,\omega_i)]$ is the fourier transform of the wave function given in \cref{equ-wave}. Thus, the spectral function $M\qty[(\omega_s-\omega_i-\omega_-)\frac{\tau}{2}]=\exp\qty[-(\omega_s-\omega_i-\omega_-)^2\tau^2/4]$ is responsible for the time-gate function $\mathcal{M}(\frac{T-t}{\tau})$. The frequency scale of the spectrum is from the term $e^{i\omega(t_4-t_3)}$ interrogated with the four-point dipole correlation $\bra{\Psi_F}\text{V}(t_1)\text{V}^{\dagger}(t_3)\text{V}(t_4)\text{V}^{\dagger}(t_2)\ket{\Psi_F}$. Detailed derivations are given in the Supporting Information.

Mingran Zhang and Yihe Xu contributed equally to this work.
\section*{Acknowledgements}

Z. D. Z., Y. X. and M. Z. gratefully  acknowledge the support of the Excellent Young Scientists Fund by National Science
Foundation of China (No. 9240172), the General Fund by National Science Foundation of China (No. 12474364), and the National Science Foundation of China/RGC Collaborative Research Scheme (No. 9054901).

\section*{Notes}
The authors declare no competing financial interest.\par

\begin{itemize}
  \item SI.pdf: Parameters in the Hamiltonian for PBI-1 trimer used in computations and some setups of TD-DMRG dynamics; Derivation of quantum time-frequency resolved spectral signals of \cref{equ-time-frequency-resolved}; Derivation of laser-based time-gated fluorescence spectrum; Experimental Feasibility and Testable Predictions.
\end{itemize}


\bibliography{refs}

@Article{1999_Mirko_JCP,
  author   = {Grimme, Stefan and Waletzke, Mirko},
  journal  = {J. Chem. Phys.},
  pages    = {5645-5655},
  title    = {A combination of Kohn-Sham density functional theory and multi-reference configuration interaction methods},
  year     = {1999},
  issn     = {0021-9606},
  month    = {10},
  number   = {13},
  volume   = {111},
  doi      = {10.1063/1.479866},
  eprint   = {https://pubs.aip.org/aip/jcp/article-pdf/111/13/5645/19112751/5645_1_online.pdf},
  url      = {https://doi.org/10.1063/1.479866},
}

@Article{2004_White_PRL,
  author    = {White, Steven R. and Feiguin, Adrian E.},
  journal   = {Phys. Rev. Lett.},
  pages     = {076401},
  title     = {Real-Time Evolution Using the Density Matrix Renormalization Group},
  year      = {2004},
  month     = {Aug},
  volume    = {93},
  doi       = {10.1103/PhysRevLett.93.076401},
  issue     = {7},
  numpages  = {4},
  publisher = {American Physical Society},
  url       = {https://link.aps.org/doi/10.1103/PhysRevLett.93.076401},
}

@Article{2011_PCCP_Oliver,
  author    = {Ambrosek, David and Marciniak, Henning and Lochbrunner, Stefan and Tatchen, Jörg and Li, Xue-Qing and Würthner, Frank and Kühn, Oliver},
  journal   = {Phys. Chem. Chem. Phys.},
  pages     = {17649-17657},
  title     = {Photophysical and quantum chemical study on a J-aggregate forming perylene bisimide monomer},
  year      = {2011},
  volume    = {13},
  doi       = {10.1039/C1CP21624D},
  issue     = {39},
  publisher = {The Royal Society of Chemistry},
  url       = {http://dx.doi.org/10.1039/C1CP21624D},
}

@Article{2011_JPCA_Stefan,
  author  = {Marciniak, Henning and Li, Xue-Qing and W{\"u}rthner, Frank and Lochbrunner, Stefan},
  journal = {J. Phys. Chem. A},
  pages   = {648-654},
  title   = {One-Dimensional Exciton Diffusion in Perylene Bisimide Aggregates},
  year    = {2011},
  number  = {5},
  volume  = {115},
  doi     = {10.1021/jp107407p},
  eprint  = { https://doi.org/10.1021/jp107407p},
  url     = {https://doi.org/10.1021/jp107407p},
}

@Article{2012_JPCA_Kuhn,
  author  = {Ambrosek, D. and K{\"o}hn, A. and Schulze, J. and K{\"u}hn, O.},
  journal = {J. Phys. Chem. A},
  pages   = {11451-11458},
  title   = {Quantum Chemical Parametrization and Spectroscopic Characterization of the Frenkel Exciton Hamiltonian for a J-Aggregate Forming Perylene Bisimide Dye},
  year    = {2012},
  number  = {46},
  volume  = {116},
  doi     = {10.1021/jp3069706},
  eprint  = {https://doi.org/10.1021/jp3069706},
  url     = {https://doi.org/10.1021/jp3069706},
}

@Article{2013_JPCA_Kuhn,
  author  = {Schr{\"o}ter, M. and K{\"u}hn, O.},
  journal = {J. Phys. Chem. A},
  pages   = {7580-7588},
  title   = {Interplay Between Nonadiabatic Dynamics and Frenkel Exciton Transfer in Molecular Aggregates: Formulation and Application to a Perylene Bismide Model},
  year    = {2013},
  number  = {32},
  volume  = {117},
  doi     = {10.1021/jp402587p},
  eprint  = {https://doi.org/10.1021/jp402587p},
  url     = {https://doi.org/10.1021/jp402587p},
}

@Article{2014_NC_Fuller,
  author  = {Fuller, Franklin D. and Pan, Jie and Gelzinis, Andrius and Butkus, Vytautas and Senlik, S. Seckin and Wilcox, Daniel E. and Yocum, Charles F. and Valkunas, Leonas and Abramavicius, Darius and Ogilvie, Jennifer P.},
  journal = {Nat. Chem.},
  pages   = {706-711},
  title   = {Vibronic coherence in oxygenic photosynthesis},
  year    = {2014},
  issn    = {1755-4349},
  month   = {Aug},
  number  = {8},
  volume  = {6},
  doi     = {10.1038/nchem.2005},
  url     = {https://doi.org/10.1038/nchem.2005},
}

@Article{2015_PR_SCHROTER,
  author           = {M. Schröter and S.D. Ivanov and J. Schulze and S.P. Polyutov and Y. Yan and T. Pullerits and O. Kühn},
  journal          = {Phys. Rep.},
  pages            = {1-78},
  title            = {Exciton–vibrational coupling in the dynamics and spectroscopy of Frenkel excitons in molecular aggregates},
  year             = {2015},
  issn             = {0370-1573},
  volume           = {567},
  doi              = {https://doi.org/10.1016/j.physrep.2014.12.001},
  modificationdate = {2026-01-22T12:19:03},
  url              = {https://www.sciencedirect.com/science/article/pii/S0370157314004104},
}

@Article{2019_PAECKEL_AP,
  author   = {Sebastian Paeckel and Thomas Köhler and Andreas Swoboda and Salvatore R. Manmana and Ulrich Schollwöck and Claudius Hubig},
  journal  = {Ann. Phys.},
  pages    = {167998},
  title    = {Time-evolution Methods for Matrix-Product States},
  year     = {2019},
  issn     = {0003-4916},
  volume   = {411},
  doi      = {https://doi.org/10.1016/j.aop.2019.167998},
  url      = {https://www.sciencedirect.com/science/article/pii/S0003491619302532},
}

@Article{2019_NC_Gaynor,
  author  = {Gaynor, James D.
and Sandwisch, Jason
and Khalil, Munira},
  journal = {Nat. Commun.},
  pages   = {5621},
  title   = {Vibronic coherence evolution in multidimensional ultrafast photochemical processes},
  year    = {2019},
  issn    = {2041-1723},
  month   = {Dec},
  number  = {1},
  volume  = {10},
  doi     = {10.1038/s41467-019-13503-9},
  url     = {https://doi.org/10.1038/s41467-019-13503-9},
}

@Article{2020_JACS_Macro,
  author  = {Conti, Irene and Cerullo, Giulio and Nenov, Artur and Garavelli, Marco},
  journal = {J. Am. Chem. Soc.},
  pages   = {16117-16139},
  title   = {Ultrafast Spectroscopy of Photoactive Molecular Systems from First Principles: Where We Stand Today and Where We Are Going},
  year    = {2020},
  number  = {38},
  volume  = {142},
  doi     = {10.1021/jacs.0c04952},
  eprint  = {https://doi.org/10.1021/jacs.0c04952},
  url     = {https://doi.org/10.1021/jacs.0c04952},
}

@Article{2022_JACS_Kim,
  author  = {Hong, Yongseok and Schlosser, Felix and Kim, Woojae and W{\"u}rthner, Frank and Kim, Dongho},
  journal = {J. Am. Chem. Soc.},
  pages   = {15539-15548},
  title   = {Ultrafast Symmetry-Breaking Charge Separation in a Perylene Bisimide Dimer Enabled by Vibronic Coupling and Breakdown of Adiabaticity},
  year    = {2022},
  number  = {34},
  volume  = {144},
  doi     = {10.1021/jacs.2c03916},
  eprint  = {https://doi.org/10.1021/jacs.2c03916},
  url     = {https://doi.org/10.1021/jacs.2c03916},
}

@Article{2022_NC_Sil,
  author  = {Sil, Sourav
and Tilluck, Ryan W.
and Mohan T. M., Nila
and Leslie, Chase H.
and Rose, Justin B.
and Dom{\'i}nguez-Mart{\'i}n, Maria Agustina
and Lou, Wenjing
and Kerfeld, Cheryl A.
and Beck, Warren F.},
  journal = {Nat. Chem.},
  pages   = {1286-1294},
  title   = {Excitation energy transfer and vibronic coherence in intact phycobilisomes},
  year    = {2022},
  issn    = {1755-4349},
  month   = {Nov},
  number  = {11},
  volume  = {14},
  doi     = {10.1038/s41557-022-01026-8},
  url     = {https://doi.org/10.1038/s41557-022-01026-8},
}

@Article{2022_SA_Policht,
  author  = {Veronica R. Policht  and Andrew Niedringhaus  and Rhiannon Willow  and Philip D. Laible  and David F. Bocian  and Christine Kirmaier  and Dewey Holten  and Tomáš Mančal  and Jennifer P. Ogilvie },
  journal = {Sci. Adv.},
  pages   = {eabk0953},
  title   = {Hidden vibronic and excitonic structure and vibronic coherence transfer in the bacterial reaction center},
  year    = {2022},
  number  = {1},
  volume  = {8},
  doi     = {10.1126/sciadv.abk0953},
  eprint  = {https://www.science.org/doi/pdf/10.1126/sciadv.abk0953},
  url     = {https://www.science.org/doi/abs/10.1126/sciadv.abk0953},
}

@Article{2022_Shuai_WCMS,
  author   = {Ren, Jiajun and Li, Weitang and Jiang, Tong and Wang, Yuanheng and Shuai, Zhigang},
  journal  = {WIREs Comput. Mol. Sci.},
  pages    = {e1614},
  title    = {Time-dependent Density Matrix Renormalization Group Method for Quantum Dynamics in Complex Systems},
  year     = {2022},
  **eprint = {https:\_wires.onlinelibrary.wiley.com/doi/pdf/10.1002/wcms.1614},
  _number  = {n/a},
  _volume  = {n/a},
  doi      = {https:doi.org/10.1002/wcms.1614},
  keywords = {density matrix renormalization group, matrix product operator, matrix product state, quantum dynamics},
  url      = {https:\_wires.onlinelibrary.wiley.com/doi/abs/10.1002/wcms.1614},
}

@Article{2023_Xu_JCTC2,
  author  = {Xu, Yihe and Cheng, Yifan and Song, Yinxuan and Ma, Haibo},
  journal = {J. Chem. Theory Comput.},
  pages   = {4781-4795},
  title   = {New Density Matrix Renormalization Group Approaches for Strongly Correlated Systems Coupled with Large Environments},
  year    = {2023},
  number  = {15},
  volume  = {19},
  doi     = {10.1021/acs.jctc.2c01316},
  eprint  = {https://doi.org/10.1021/acs.jctc.2c01316},
  url     = {https://doi.org/10.1021/acs.jctc.2c01316},
}

@Article{2024_Frank_JACS,
  author  = {Sukumaran, Divya P. and Shoyama, Kazutaka and Dubey, Rajeev K. and W{\"u}rthner, Frank},
  journal = {J. Am. Chem. Soc.},
  pages   = {22077-22084},
  title   = {Cooperative Binding and Chirogenesis in an Expanded Perylene Bisimide Cyclophane},
  year    = {2024},
  number  = {31},
  volume  = {146},
  doi     = {10.1021/jacs.4c08073},
  eprint  = {https://doi.org/10.1021/jacs.4c08073},
  url     = {https://doi.org/10.1021/jacs.4c08073},
}

@Article{2025_NChem_Ernst,
  author    = {Ernst, Leander and Song, Hongwei and Kim, Dongho and W{\"u}rthner, Frank},
  journal   = {Nat. Chem.},
  pages     = {1--10},
  title     = {Photoinduced stepwise charge hopping in $\pi$-stacked perylene bisimide donor--bridge--acceptor arrays},
  year      = {2025},
  publisher = {Nature Publishing Group UK London},
}

@Article{2025_Angew_Frank,
  author   = {Teichmann, Ben and Sárosi, Menyhárt and Shoyama, Kazutaka and Niyas, M. A. and Dubey, Rajeev K. and Würthner, Frank},
  journal  = {Angew. Chem., Int. Ed.},
  pages    = {e202414069},
  title    = {‘Invisible’ Molecular Dynamics Revealed for a Conformationally Chiral $\pi$-Stacked Perylene Bisimide Foldamer},
  year     = {2025},
  number   = {2},
  volume   = {64},
  doi      = {https://doi.org/10.1002/anie.202414069},
  eprint   = {https://onlinelibrary.wiley.com/doi/pdf/10.1002/anie.202414069},
  url      = {https://onlinelibrary.wiley.com/doi/abs/10.1002/anie.202414069},
}

@Article{15-S-Universallinearoptics,
  author           = {J. Carolan and C. Harrold and C. Sparrow and E. Martin-Lopez and N. J. Russell and J. W. Silverstone and P. J. Shadbolt and N. Matsuda and M. Oguma and M. Itoh and G. D. Marshall and M. G. Thompson and J. C. F. Matthews and T. Hashimoto and J. L. O{\textquotesingle}Brien and A. Laing},
  journal          = {Science},
  pages            = {711--716},
  title            = {Universal linear optics},
  year             = {2015},
  month            = {jul},
  number           = {6249},
  volume           = {349},
  comment          = {六个模式的光子干涉仪道路，利用了30个热光相位调制器的集成电路，允许对六个通道的动态调谐达到所有任意的六模式线性干涉仪。
展示了很多量子线路（例如纠缠态、CNOT门、旋转门等）的实验方案，说明这六个模式能够实现相当多的功能，同时对玻色采样也进行了一个验证。
提出了一种通用线性光学处理器（LPU）的设想，},
  doi              = {10.1126/science.aab3642},
  modificationdate = {2022-02-16T17:04:07},
  publisher        = {American Association for the Advancement of Science ({AAAS})},
  ranking          = {rank5},
  readstatus       = {read},
  relevance        = {relevant},
}

@Article{25-S-classicalcomputationquantum,
  author               = {King, Andrew D. and Nocera, Alberto and Rams, Marek M. and Dziarmaga, Jacek and Wiersema, Roeland and Bernoudy, William and Raymond, Jack and Kaushal, Nitin and Heinsdorf, Niclas and Harris, Richard and Boothby, Kelly and Altomare, Fabio and Asad, Mohsen and Berkley, Andrew J. and Boschnak, Martin and Chern, Kevin and Christiani, Holly and Cibere, Samantha and Connor, Jake and Dehn, Martin H. and Deshpande, Rahul and Ejtemaee, Sara and Farre, Pau and Hamer, Kelsey and Hoskinson, Emile and Huang, Shuiyuan and Johnson, Mark W. and Kortas, Samuel and Ladizinsky, Eric and Lanting, Trevor and Lai, Tony and Li, Ryan and MacDonald, Allison J. R. and Marsden, Gaelen and McGeoch, Catherine C. and Molavi, Reza and Oh, Travis and Neufeld, Richard and Norouzpour, Mana and Pasvolsky, Joel and Poitras, Patrick and Prescott, Thomas and Reis, Mauricio and Rich, Chris and Samani, Mohammad and Sheldan, Benjamin and Smirnov, Anatoly and Sterpka, Edward and Trullas Clavera, Berta and Tsai, Nicholas and Volkmann, Mark and Whiticar, Alexander M. and Whittaker, Jed D. and Wilkinson, Warren and Yao, Jason and Yi, T. J. and Sandvik, Anders W. and Alvarez, Gonzalo and Melko, Roger G. and Carrasquilla, Juan and Franz, Marcel and Amin, Mohammad H.},
  journal              = {Science},
  pages                = {199-204},
  title                = {Beyond-classical computation in quantum simulation},
  year                 = {2025},
  issn                 = {0036-8075},
  month                = {APR 10},
  number               = {6743},
  volume               = {388},
  doi                  = {10.1126/science.ado6285},
  eissn                = {1095-9203},
  orcid-numbers        = {Ra, Marek/0000-0002-1235-7758 Poitras, Patrick/0000-0002-9341-4389},
  researcherid-numbers = {Nocera, Alberto/A-6329-2016 Rams, Marek/E-1598-2016 Ra, Marek/E-1598-2016 Whiticar, Alexander/L-8291-2016 Kaushal, Nitin/HPD-4393-2023},
  unique-id            = {WOS:001494298000035},
}

@Article{25-PRA-Generationarbitraryhigh,
  author               = {Du, Fang-Fang and Ma, Ming and Bai, Zhuo-Ya and Tan, Qiu-lin},
  journal              = {Phys. Rev. A},
  pages                = {032604},
  title                = {Generation of arbitrary high-dimensional qudit-based entangled states},
  year                 = {2025},
  issn                 = {2469-9934},
  month                = {MAR 6},
  number               = {3},
  volume               = {111},
  article-number       = {032604},
  doi                  = {10.1103/PhysRevA.111.032604},
  eissn                = {2469-9934},
  modificationdate     = {2026-01-22T11:49:28},
  publisher            = {American Physical Society (APS)},
  researcherid-numbers = {Bai, Zhuoya/AAH-2943-2020},
  unique-id            = {WOS:001447534300001},
}

@Article{25-SA-Simultaneoustransmissioninformation,
  author               = {Pan, Dong and Liu, Yu-Chen and Niu, Penghao and Zhang, Haoran and Zhang, Feihao and Wang, Min and Song, Xiao-Tian and Chen, Xiuwei and Zheng, Chao and Long, Gui-Lu},
  journal              = {Sci. Adv.},
  pages                = {eadt4627},
  title                = {Simultaneous transmission of information and key exchange using the same photonic quantum states},
  year                 = {2025},
  issn                 = {2375-2548},
  month                = {FEB 21},
  number               = {8},
  volume               = {11},
  article-number       = {eadt4627},
  doi                  = {10.1126/sciadv.adt4627},
  modificationdate     = {2026-01-22T11:51:33},
  orcid-numbers        = {Pan, Dong/0000-0002-3922-1308 Long, Gui Lu/0000-0002-9023-1579 Wang, Min/0000-0001-5321-3323 Liu, Yu-chen/0009-0002-9296-6752 CHEN, XIUWEI/0000-0003-0756-504X Zheng, Chao/0000-0002-8942-6651 Zhang, Haoran/0000-0001-6261-038X},
  researcherid-numbers = {Pan, Dong/HJG-5632-2022 Long, Gui Lu/B-1170-2008 Zheng, Chao/JNT-2046-2023},
  unique-id            = {WOS:001428018300003},
}

@Article{25-NP-Measuringquantumstate,
  author               = {Laurell, Hugo and Luo, Sizuo and Weissenbilder, Robin and Ammitzboell, Mattias and Ahmed, Shahnawaz and Soederberg, Hugo and Petersson, C. Leon. M. and Poulain, Venus and Guo, Chen and Dittel, Christoph and Finkelstein-Shapiro, Daniel and Squibb, Richard J. and Feifel, Raimund and Gisselbrecht, Mathieu and Arnold, Cord L. and Buchleitner, Andreas and Lindroth, Eva and Frisk Kockum, Anton and L'Huillier, Anne and Busto, David},
  journal              = {Nat. Photonics},
  pages                = {352-357},
  title                = {Measuring the quantum state of photoelectrons},
  year                 = {2025},
  issn                 = {1749-4885},
  month                = {APR},
  number               = {4},
  volume               = {19},
  doi                  = {10.1038/s41566-024-01607-8},
  earlyaccessdate      = {JAN 2025},
  eissn                = {1749-4893},
  orcid-numbers        = {Guo, Chen/0000-0003-3532-143X Busto Ortiz, David/0000-0003-4311-3315 Linoth, Eva/0000-0003-3444-1317 Laurell, Hugo/0000-0002-2059-2715 L'Huillier, Anne/0000-0002-1335-4022 Weissenbilder, Robin/0000-0002-8081-3727},
  researcherid-numbers = {Arnold, Cord/B-9365-2014 Kockum, Anton/E-8255-2014 Busto, David/L-7679-2018 Dittel, Christoph/B-4636-2017 Busto Ortiz, David/L-7679-2018 Linoth, Eva/I-2715-2013 L'Huillier, Anne/P-4379-2015},
  unique-id            = {WOS:001408532900001},
}

@Article{25-PRA-Quantumenhanceddistributed,
  author    = {Zhang, Zhongjian and Wang, Ben and Zheng, Kaimin and Zhang, Lijian},
  journal   = {Phys. Rev. A},
  pages     = {062403},
  title     = {Quantum-enhanced distributed sensing using nonlinear optical networks},
  year      = {2025},
  month     = {Dec},
  volume    = {112},
  doi       = {10.1103/p77r-f5kf},
  issue     = {6},
  numpages  = {13},
  publisher = {American Physical Society},
  url       = {https://link.aps.org/doi/10.1103/p77r-f5kf},
}

@Article{25-PRL-ApproachingMultiparameterQuantum,
  author           = {Mi, Minghao and Wang, Ben and Zhang, Lijian},
  journal          = {Phys. Rev. Lett.},
  pages            = {110804},
  title            = {Approaching the Multiparameter Quantum Cram\'er-Rao Bound via Classical Correlation and Entangling Measurements},
  year             = {2025},
  month            = {Sep},
  volume           = {135},
  doi              = {10.1103/37kb-qq6k},
  issue            = {11},
  modificationdate = {2026-01-01T15:02:00},
  numpages         = {7},
  publisher        = {American Physical Society},
  url              = {https://link.aps.org/doi/10.1103/37kb-qq6k},
}

@Article{23-N-Singlephotonabsorption,
  author           = {Li, Quanwei and Orcutt, Kaydren and Cook, Robert L. and Sabines-Chesterking, Javier and Tong, Ashley L. and Schlau-Cohen, Gabriela S. and Zhang, Xiang and Fleming, Graham R. and Whaley, K. Birgitta},
  journal          = {Nature},
  pages            = {300--304},
  title            = {Single-photon absorption and emission from a natural photosynthetic complex},
  year             = {2023},
  issn             = {1476-4687},
  month            = {jun},
  number           = {7969},
  volume           = {619},
  creationdate     = {2026-01-01T15:17:31},
  doi              = {10.1038/s41586-023-06121-5},
  modificationdate = {2026-01-01T15:17:39},
  publisher        = {Springer Science and Business Media LLC},
}

@Article{14-TJoPCL-StimulatedRamanSpectroscopy,
  author           = {Dorfman, Konstantin E. and Schlawin, Frank and Mukamel, Shaul},
  journal          = {J. Phys. Chem. Lett.},
  pages            = {2843--2849},
  title            = {Stimulated Raman Spectroscopy with Entangled Light: Enhanced Resolution and Pathway Selection},
  year             = {2014},
  issn             = {1948-7185},
  month            = {aug},
  number           = {16},
  volume           = {5},
  creationdate     = {2026-01-18T13:40:40},
  doi              = {10.1021/jz501124a},
  modificationdate = {2026-01-18T13:41:03},
  publisher        = {American Chemical Society (ACS)},
}

@Article{09-PRA-Nonlinearspectroscopyentangled,
  author           = {Roslyak, Oleksiy and Marx, Christoph A. and Mukamel, Shaul},
  journal          = {Phys. Rev. A},
  pages            = {033832},
  title            = {Nonlinear spectroscopy with entangled photons: Manipulating quantum pathways of matter},
  year             = {2009},
  issn             = {1094-1622},
  month            = {mar},
  number           = {3},
  volume           = {79},
  creationdate     = {2026-01-18T13:41:52},
  doi              = {10.1103/physreva.79.033832},
  modificationdate = {2026-01-18T13:42:26},
  publisher        = {American Physical Society (APS)},
}

@Article{08-PRA-Nonlinearopticalspectroscopy,
  author           = {Marx, Christoph A. and Harbola, Upendra and Mukamel, Shaul},
  journal          = {Phys. Rev. A},
  pages            = {022110},
  title            = {Nonlinear optical spectroscopy of single, few, and many molecules: Nonequilibrium Green’s function QED approach},
  year             = {2008},
  issn             = {1094-1622},
  month            = {feb},
  number           = {2},
  volume           = {77},
  creationdate     = {2026-01-18T13:46:30},
  doi              = {10.1103/PhysRevA.77.022110},
  modificationdate = {2026-01-18T13:46:38},
  publisher        = {American Physical Society (APS)},
}

@Article{13-NJoP-Rolespectralshape,
  author           = {de J León-Montiel, R and Svozilík, J and Salazar-Serrano, L J and Torres, Juan P},
  journal          = {New J. Phys.},
  pages            = {053023},
  title            = {Role of the spectral shape of quantum correlations in two-photon virtual-state spectroscopy},
  year             = {2013},
  issn             = {1367-2630},
  month            = {may},
  number           = {5},
  volume           = {15},
  creationdate     = {2026-01-18T13:47:56},
  doi              = {10.1088/1367-2630/15/5/053023},
  modificationdate = {2026-01-18T13:48:03},
  publisher        = {IOP Publishing},
}

@Article{97-PRA-Theorytwophoton,
  author           = {Keller, Timothy E. and Rubin, Morton H.},
  journal          = {Phys. Rev. A},
  pages            = {1534--1541},
  title            = {Theory of two-photon entanglement for spontaneous parametric down-conversion driven by a narrow pump pulse},
  year             = {1997},
  issn             = {1094-1622},
  month            = {aug},
  number           = {2},
  volume           = {56},
  creationdate     = {2026-01-18T13:51:26},
  doi              = {10.1103/PhysRevA.56.1534},
  modificationdate = {2026-01-18T13:51:40},
  publisher        = {American Physical Society (APS)},
}

@Article{10-PR-Spatialcorrelationsparametric,
  author           = {Walborn, S.P. and Monken, C.H. and Pádua, S. and Souto Ribeiro, P.H.},
  journal          = {Phys. Rep.},
  pages            = {87--139},
  title            = {Spatial correlations in parametric down-conversion},
  year             = {2010},
  issn             = {0370-1573},
  month            = {oct},
  number           = {4–5},
  volume           = {495},
  creationdate     = {2026-01-18T13:52:39},
  doi              = {10.1016/j.physrep.2010.06.003},
  modificationdate = {2026-01-18T13:52:46},
  publisher        = {Elsevier BV},
}

@Article{16-PS-Timefrequencygated,
  author           = {Dorfman, Konstantin E and Mukamel, Shaul},
  journal          = {Phys. Scr.},
  pages            = {083004},
  title            = {Time-and-frequency-gated photon coincidence counting; a novel multidimensional spectroscopy tool},
  year             = {2016},
  issn             = {1402-4896},
  month            = {jul},
  number           = {8},
  volume           = {91},
  creationdate     = {2026-01-18T13:56:06},
  doi              = {10.1088/0031-8949/91/8/083004},
  modificationdate = {2026-01-18T13:56:13},
  publisher        = {IOP Publishing},
}

@Misc{25--Twodimensionalfluorescence,
  author           = {Yuta Fujihashi and Akihito Ishizaki},
  title            = {Two-dimensional fluorescence spectroscopy with quantum entangled photons: Idler-referenced timing without pump detection},
  year             = {2025},
  archiveprefix    = {arXiv},
  eprint           = {2511.22198},
  modificationdate = {2026-05-30T16:09:24},
  primaryclass     = {physics.chem-ph},
  url              = {https://arxiv.org/abs/2511.22198},
}

@Article{21-NC-Quantumenhancedmultiple,
  author           = {Seongjin Hong and Junaid ur Rehman and Yong-Su Kim and Young-Wook Cho and Seung-Woo Lee and Hojoong Jung and Sung Moon and Sang-Wook Han and Hyang-Tag Lim},
  journal          = {Nat. Commun.},
  pages            = {5211},
  title            = {Quantum enhanced multiple-phase estimation with multi-mode N00N states},
  year             = {2021},
  issn             = {2041-1723},
  month            = {sep},
  number           = {1},
  volume           = {12},
  comment          = {21 第一个N00N态的多相位估计
    Quantum enhanced multiple-phase estimation with multi-mode N00N states
        利用SPDC产生光子对，同时采用偏振光来制备多模2002态，通过PNRD组产生伪光子计数探测符合计数事件，最后使用MLE来估计参数
        实验对比了采用单光子FOCK态（即19op）和不对称NOON态（即QCRB取得最小值的态，见13 多模N00N态可以达到更高精度理论）的情况，说明该实验比FOCK态精度更高，虽然理论精度上限不如13，但是其实际可达到精度（CRB）比13更高。实验达到误差为1.85，比Fock态的2.44要低，同时高于HL（1.5）。},
  creationdate     = {2026-01-22T11:53:25},
  doi              = {10.1038/s41467-021-25451-4},
  modificationdate = {2026-01-22T11:53:25},
  publisher        = {Springer Science and Business Media {LLC}},
  refid            = {Hong2021},
  url              = {https://doi.org/10.1038/s41467-021-25451-4},
}

@Article{23-JCP-Probingexcitondynamics,
  author           = {Fujihashi, Yuta and Miwa, Kuniyuki and Higashi, Masahiro and Ishizaki, Akihito},
  journal          = {J. Chem. Phys.},
  pages            = {114201},
  title            = {Probing exciton dynamics with spectral selectivity through the use of quantum entangled photons},
  year             = {2023},
  issn             = {1089-7690},
  month            = {sep},
  number           = {11},
  volume           = {159},
  creationdate     = {2026-01-22T12:00:57},
  doi              = {10.1063/5.0169768},
  eprint           = {https://pubs.aip.org/aip/jcp/article-pdf/doi/10.1063/5.0169768/18126987/114201_1_5.0169768.pdf},
  modificationdate = {2026-01-22T12:00:57},
  publisher        = {AIP Publishing},
  url              = {https://doi.org/10.1063/5.0169768},
}

@Article{25-NC-Correlatedphotontime,
  author           = {Álvarez-Mendoza, Raúl and Uboldi, Lorenzo and Lyons, Ashley and Cogdell, Richard J. and Cerullo, Giulio and Faccio, Daniele},
  journal          = {Nat. Commun.},
  pages            = {8634},
  title            = {Correlated-photon time- and frequency-resolved optical spectroscopy},
  year             = {2025},
  issn             = {2041-1723},
  month            = {sep},
  number           = {1},
  volume           = {16},
  creationdate     = {2026-01-22T12:02:11},
  doi              = {10.1038/s41467-025-63830-3},
  modificationdate = {2026-01-22T12:02:11},
  publisher        = {Springer Science and Business Media LLC},
  refid            = {Álvarez-Mendoza2025},
  url              = {https://doi.org/10.1038/s41467-025-63830-3},
}

@Article{13-NC-Suppressionpopulationtransport,
  author           = {Schlawin, Frank and Dorfman, Konstantin E. and Fingerhut, Benjamin P. and Mukamel, Shaul},
  journal          = {Nat. Commun.},
  pages            = {1782},
  title            = {Suppression of population transport and control of exciton distributions by entangled photons},
  year             = {2013},
  issn             = {2041-1723},
  month            = {apr},
  number           = {1},
  volume           = {4},
  creationdate     = {2026-01-22T12:03:45},
  doi              = {10.1038/ncomms2802},
  modificationdate = {2026-01-22T12:03:45},
  publisher        = {Springer Science and Business Media LLC},
  refid            = {Schlawin2013},
  url              = {https://doi.org/10.1038/ncomms2802},
}

@Article{2024_Xu_JCP,
  author   = {Xu, Yihe and Liu, Chungen and Ma, Haibo},
  journal  = {J. Chem. Phys.},
  pages    = {052501},
  title    = {Kylin-V: An open-source package calculating the dynamic and spectroscopic properties of large systems},
  year     = {2024},
  issn     = {0021-9606},
  month    = {08},
  number   = {5},
  volume   = {161},
  doi      = {10.1063/5.0220712},
  eprint   = {https://pubs.aip.org/aip/jcp/article-pdf/doi/10.1063/5.0220712/20090403/052501_1_5.0220712.pdf},
  url      = {https://doi.org/10.1063/5.0220712},
}

@Article{2025_cy_JACS,
  author  = {Burigana, Vittoria and Buttarazzi, Edoardo and Toffoletti, Federico and Fresch, Elisa and Tumbarello, Francesco and Petrone, Alessio and Collini, Elisabetta},
  journal = {J. Am. Chem. Soc.},
  pages   = {32994-33002},
  title   = {Signatures of a Conical Intersection in Two-Dimensional Spectra of a Red-Absorbing Squaraine Dye},
  year    = {2025},
  number  = {36},
  volume  = {147},
  doi     = {10.1021/jacs.5c10393},
  eprint  = {https://doi.org/10.1021/jacs.5c10393},
  url     = {https://doi.org/10.1021/jacs.5c10393},
}

@Article{Liu2025,
  author  = {Liu, Diankai and He, Zixu and Gao, Wenjie and Shang, Jizhen and Yang, Yiqing and Zhang, Xiaofan and Li, Xiaohua and Ma, Huimin and Shi, Wen},
  journal = {Nat. Commun.},
  pages   = {4911},
  title   = {Near-infrared II cyanine fluorophores with large stokes shift engineered by regulating respective absorption and emission},
  year    = {2025},
  month   = {May},
  number  = {1},
  volume  = {16},
  day     = {27},
  doi     = {10.1038/s41467-025-60241-2},
  ssn     = {2041-1723},
  url     = {https://doi.org/10.1038/s41467-025-60241-2},
}

@Article{2024_acc_res_cy,
  author  = {Zhao, Xueze and Du, Jianjun and Sun, Wen and Fan, Jiangli and Peng, Xiaojun},
  journal = {Acc. Chem. Res},
  pages   = {2582-2593},
  title   = {Regulating Charge Transfer in Cyanine Dyes: A Universal Methodology for Enhancing Cancer Phototherapeutic Efficacy},
  year    = {2024},
  number  = {17},
  volume  = {57},
  doi     = {10.1021/acs.accounts.4c00399},
  eprint  = {https://doi.org/10.1021/acs.accounts.4c00399},
  url     = {https://doi.org/10.1021/acs.accounts.4c00399},
}

@Article{2024_ry_JACS,
  author  = {Wang, Kangwei and You, Xiaoxiao and Miao, Xiaodan and Yi, Yuanping and Peng, Shaoqian and Wu, Di and Chen, Xingyu and Xu, Jingwen and Sfeir, Matthew Y. and Xia, Jianlong},
  journal = {J. Am. Chem. Soc.},
  pages   = {13326-13335},
  title   = {Activated Singlet Fission Dictated by Anti-Kasha Property in a Rylene Imide Dye},
  year    = {2024},
  number  = {19},
  volume  = {146},
  doi     = {10.1021/jacs.4c01732},
  eprint  = {https://doi.org/10.1021/jacs.4c01732},
  url     = {https://doi.org/10.1021/jacs.4c01732},
}

@Article{2024_poly_pccp,
  author    = {Obloy, Laura M. and Jockusch, Steffen and Tarnovsky, Alexander N.},
  journal   = {Phys. Chem. Chem. Phys.},
  pages     = {24261-24278},
  title     = {Shortwave infrared polymethine dyes for bioimaging: ultrafast relaxation dynamics and excited-state decay pathways},
  year      = {2024},
  volume    = {26},
  doi       = {10.1039/D4CP01411A},
  issue     = {37},
  publisher = {The Royal Society of Chemistry},
  url       = {http://dx.doi.org/10.1039/D4CP01411A},
}

@Article{2025_JPCL_Eric,
  author  = {Cruz Neto, Daniel H. and Sucre-Rosales, Estefanía and Vauthey, Eric},
  journal = {J. Phys. Chem. Lett.},
  pages   = {13241-13246},
  title   = {Making Charge Recombination Spin-Forbidden for Efficient Generation and Excitation of Perylene Diimide Radical Anion by Pump–Pump–Probe Spectroscopy},
  year    = {2025},
  number  = {51},
  volume  = {16},
  doi     = {10.1021/acs.jpclett.5c03240},
  eprint  = { https://doi.org/10.1021/acs.jpclett.5c03240},
  url     = {https://doi.org/10.1021/acs.jpclett.5c03240},
}

@Article{2024_JCP_James,
  author  = {O’Connor, James P. and Schultz, Jonathan D. and Tcyrulnikov, Nikolai A. and Kim, Taeyeon and Young, Ryan M. and Wasielewski, Michael R.},
  journal = {J. Chem. Phys.},
  pages   = {074306},
  title   = {Distinct vibrational motions promote disparate excited-state decay pathways in cofacial perylenediimide dimers},
  year    = {2024},
  issn    = {0021-9606},
  month   = {08},
  number  = {7},
  volume  = {161},
  doi     = {10.1063/5.0218752},
  eprint  = {https://pubs.aip.org/aip/jcp/article-pdf/doi/10.1063/5.0218752/20113007/074306_1_5.0218752.pdf},
  url     = {https://doi.org/10.1063/5.0218752},
}

@Article{2025_Sotome_JCP,
  author  = {Sotome, Hikaru and Higashi, Masahiro and Tanaka, Yuki and Shinokubo, Hiroshi and Kobori, Yasuhiro and Fukui, Norihito},
  journal = {J. Chem. Phys.},
  pages   = {114305},
  title   = {Effect of structural bending on the photophysical properties of perylene bisimide},
  year    = {2025},
  issn    = {0021-9606},
  month   = {03},
  number  = {11},
  volume  = {162},
  doi     = {10.1063/5.0255756},
  eprint  = {https://pubs.aip.org/aip/jcp/article-pdf/doi/10.1063/5.0255756/20445974/114305_1_5.0255756.pdf},
  url     = {https://doi.org/10.1063/5.0255756},
}

@Article{26-CSR-Quantumcoherentdynamics,
  author           = {Jha, Ajay and Zheng, Fulu and Liu, Zihui and Mukamel, Shaul and Thorwart, Michael and Miller, R. J. Dwayne and Duan, Hong-Guang},
  journal          = {Chemical Society Reviews},
  pages            = {1089--1130},
  title            = {Quantum coherent dynamics in photosynthetic protein complexes},
  year             = {2026},
  issn             = {1460-4744},
  number           = {2},
  volume           = {55},
  creationdate     = {2026-03-09T13:11:14},
  doi              = {10.1039/d5cs00948k},
  modificationdate = {2026-03-09T13:11:42},
  publisher        = {Royal Society of Chemistry (RSC)},
}

@Article{25-JotACS-AtomicTrajectoriesBimolecular,
  author           = {Xian, R. Patrick and Hayes, Stuart A. and Corthey, Gastón and Morrison, Carole A. and Marx, Alexander and Daoud, Hazem and Lu, Cheng and Miller, R. J. Dwayne},
  journal          = {Journal of the American Chemical Society},
  pages            = {28973--28980},
  title            = {Atomic Trajectories of a Bimolecular Reaction Visualized by Ultrafast Electron Diffraction},
  year             = {2025},
  issn             = {1520-5126},
  month            = jul,
  number           = {32},
  volume           = {147},
  creationdate     = {2026-03-09T13:15:23},
  doi              = {10.1021/jacs.5c07077},
  modificationdate = {2026-03-09T13:15:37},
  publisher        = {American Chemical Society (ACS)},
}

@Article{25-TJoCP-Rectificationvibrationalenergy,
  author           = {Feng, Jichen and Abraham, Ethan and Subotnik, Joseph E. and Nitzan, Abraham},
  journal          = {The Journal of Chemical Physics},
  title            = {Rectification of vibrational energy transfer in driven chiral molecules},
  year             = {2025},
  issn             = {1089-7690},
  month            = dec,
  number           = {23},
  volume           = {163},
  creationdate     = {2026-03-09T13:17:31},
  doi              = {10.1063/5.0299947},
  modificationdate = {2026-03-09T13:17:37},
  publisher        = {AIP Publishing},
}

@Article{26-CPR-phasespaceway,
  author           = {Bian, Xuezhi and Duston, Titouan and Bradbury, Nadine and Tao, Zhen and Bhati, Mansi and Qiu, Tian and Wu, Xinchun and Wu, Yanze and Subotnik, Joseph E.},
  journal          = {Chemical Physics Reviews},
  title            = {The phase-space way to electronic structure theory and subsequently chemical dynamics},
  year             = {2026},
  issn             = {2688-4070},
  month            = jan,
  number           = {1},
  volume           = {7},
  creationdate     = {2026-03-09T13:18:14},
  doi              = {10.1063/5.0286240},
  modificationdate = {2026-03-09T13:18:22},
  publisher        = {AIP Publishing},
}

@Article{17-NP-Unconditionalviolationshot,
  author           = {Slussarenko, Sergei and Weston, Morgan M. and Chrzanowski, Helen M. and Shalm, Lynden K. and Verma, Varun B. and Nam, Sae Woo and Pryde, Geoff J.},
  journal          = {Nature Photonics},
  pages            = {700--703},
  title            = {Unconditional violation of the shot-noise limit in photonic quantum metrology},
  year             = {2017},
  issn             = {1749-4893},
  month            = oct,
  number           = {11},
  volume           = {11},
  comment          = {shot noise limit相关},
  creationdate     = {2026-03-30T19:20:32},
  doi              = {10.1038/s41566-017-0011-5},
  modificationdate = {2026-03-30T19:21:01},
  publisher        = {Springer Science and Business Media LLC},
}

@Article{15-AP-PlasmonicTraceSensing,
  author           = {Pooser, Raphael C. and Lawrie, Benjamin},
  journal          = {ACS Photonics},
  pages            = {8--13},
  title            = {Plasmonic Trace Sensing below the Photon Shot Noise Limit},
  year             = {2015},
  issn             = {2330-4022},
  month            = dec,
  number           = {1},
  volume           = {3},
  comment          = {shot-noise-limit 相关},
  creationdate     = {2026-03-30T19:22:08},
  doi              = {10.1021/acsphotonics.5b00501},
  modificationdate = {2026-03-30T19:22:24},
  publisher        = {American Chemical Society (ACS)},
}

@Article{15-JotEOSP-Optimizedsignalnoise,
  author           = {Moester, M. J. B. and Ariese, F. and de Boer, J. F.},
  journal          = {Journal of the European Optical Society-Rapid Publications},
  pages            = {15022},
  title            = {Optimized signal-to-noise ratio with shot noise limited detection in Stimulated Raman Scattering microscopy},
  year             = {2015},
  issn             = {1990-2573},
  volume           = {10},
  comment          = {SNL相关},
  creationdate     = {2026-03-30T19:24:59},
  doi              = {10.2971/jeos.2015.15022},
  modificationdate = {2026-03-30T19:25:13},
  publisher        = {EDP Sciences},
}

@Article{20-TJoPCB-ShotNoiseLimited,
  author           = {Choi, Youngjin and Lim, Sohee and Shim, Joong Won and Chon, Bonghwan and Lim, Jong Min and Cho, Minhaeng},
  journal          = {The Journal of Physical Chemistry B},
  pages            = {2591--2599},
  title            = {Shot-Noise-Limited Two-Color Stimulated Raman Scattering Microscopy with a Balanced Detection Scheme},
  year             = {2020},
  issn             = {1520-5207},
  month            = mar,
  number           = {13},
  volume           = {124},
  comment          = {受激拉曼散射 (SRS) 显微镜被认为是一种通过选择性地靶向化学结构的振动模式来研究化学成分的有效技术。然而，将其应用于复杂生物环境中分子结构和动力学的观察，需要具备高分辨率和高信噪比的宽光谱覆盖范围。本文展示了一种采用平衡检测方案和光谱聚焦方法的双色 SRS 显微镜。该方法利用垂直偏振的泵浦光和斯托克斯光脉冲对产生两种不同的 SRS 信号，其中每种信号都作为另一种信号的强度参考，即使使用基于光纤的飞秒激光系统，也能显著降低背景噪声水平，使其接近散粒噪声极限。光谱聚焦方法实现了与自发拉曼散射光谱相当的高光谱分辨率。我们获得了聚合物微珠混合物以及 U2OS 细胞中脂质和蛋白质分布的双色 SRS 图像。},
  creationdate     = {2026-03-30T19:27:46},
  doi              = {10.1021/acs.jpcb.0c01065},
  modificationdate = {2026-03-30T19:28:05},
  publisher        = {American Chemical Society (ACS)},
}

@Article{23-JotACS-GeneralStrategyImprove,
  author           = {Zhang, Yuan and Yang, Chen and Peng, Sijia and Ling, Jing and Chen, Peng and Ma, Yumiao and Wang, Wenjuan and Chen, Zhixing and Chen, Chunlai},
  journal          = {Journal of the American Chemical Society},
  pages            = {4187--4198},
  title            = {General Strategy To Improve the Photon Budget of Thiol-Conjugated Cyanine Dyes},
  year             = {2023},
  issn             = {1520-5126},
  month            = feb,
  number           = {7},
  volume           = {145},
  comment          = {photobleaching
jacs},
  creationdate     = {2026-03-30T19:55:27},
  doi              = {10.1021/jacs.2c12635},
  modificationdate = {2026-03-30T19:55:50},
  publisher        = {American Chemical Society (ACS)},
}

@Article{14-O-Noninvasivenonlinearfocusing,
  author           = {Katz, Ori and Small, Eran and Guan, Yefeng and Silberberg, Yaron},
  journal          = {Optica},
  pages            = {170},
  title            = {Noninvasive nonlinear focusing and imaging through strongly scattering turbid layers},
  year             = {2014},
  issn             = {2334-2536},
  month            = sep,
  number           = {3},
  volume           = {1},
  creationdate     = {2026-03-30T20:01:20},
  doi              = {10.1364/OPTICA.1.000170},
  modificationdate = {2026-03-30T20:01:30},
  publisher        = {Optica Publishing Group},
}

@Article{20-MaAiF-Photobleachingorganicfluorophores,
  author           = {Demchenko, Alexander P},
  journal          = {Methods and Applications in Fluorescence},
  pages            = {022001},
  title            = {Photobleaching of organic fluorophores: quantitative characterization, mechanisms, protection *},
  year             = {2020},
  issn             = {2050-6120},
  month            = feb,
  number           = {2},
  volume           = {8},
  comment          = {photobleaching},
  creationdate     = {2026-03-30T20:05:43},
  doi              = {10.1088/2050-6120/ab7365},
  modificationdate = {2026-03-30T20:05:57},
  publisher        = {IOP Publishing},
}

@Article{Eshun2022,
  author    = {Eshun, Audrey and Varnavski, Oleg and Villabona-Monsalve, Juan P. and Burdick, Ryan K. and Goodson, Theodore},
  journal   = {Accounts of Chemical Research},
  pages     = {991--1003},
  title     = {Entangled Photon Spectroscopy},
  year      = {2022},
  issn      = {1520-4898},
  month     = mar,
  number    = {7},
  volume    = {55},
  comment   = {ETPA光谱},
  doi       = {10.1021/acs.accounts.1c00687},
  publisher = {American Chemical Society (ACS)},
}

@Article{Velusamy2009,
  author    = {Velusamy, Marappan and Shen, Jiun‐Yi and Lin, Jiann T. and Lin, Yi‐Chih and Hsieh, Cheng‐Chih and Lai, Chin‐Hung and Lai, Chih‐Wei and Ho, Mei‐Lin and Chen, Yu‐Chun and Chou, Pi‐Tai and Hsiao, Jong‐Kai},
  journal   = {Advanced Functional Materials},
  pages     = {2388--2397},
  title     = {A New Series of Quadrupolar Type Two‐Photon Absorption Chromophores Bearing 11, 12‐Dibutoxydibenzo[a,c]‐phenazine Bridged Amines; Their Applications in Two‐Photon Fluorescence Imaging and Two‐Photon Photodynamic Therapy},
  year      = {2009},
  issn      = {1616-3028},
  month     = aug,
  number    = {15},
  volume    = {19},
  doi       = {10.1002/adfm.200900125},
  publisher = {Wiley},
}

@Article{Li2012,
  author    = {Li, Jing‐Liang and Bao, Hong‐Chun and Hou, Xue‐Liang and Sun, Lu and Wang, Xun‐Gai and Gu, Min},
  journal   = {Angewandte Chemie International Edition},
  pages     = {1830--1834},
  title     = {Graphene Oxide Nanoparticles as a Nonbleaching Optical Probe for Two‐Photon Luminescence Imaging and Cell Therapy},
  year      = {2012},
  issn      = {1521-3773},
  month     = jan,
  number    = {8},
  volume    = {51},
  doi       = {10.1002/anie.201106102},
  publisher = {Wiley},
}

\newpage

\rule{0.05in}{1.75in}%
\begin{minipage}[b][1.75in]{3.25in}
  \sffamily
  \frenchspacing
\centering
\includegraphics[width=1.0\linewidth]{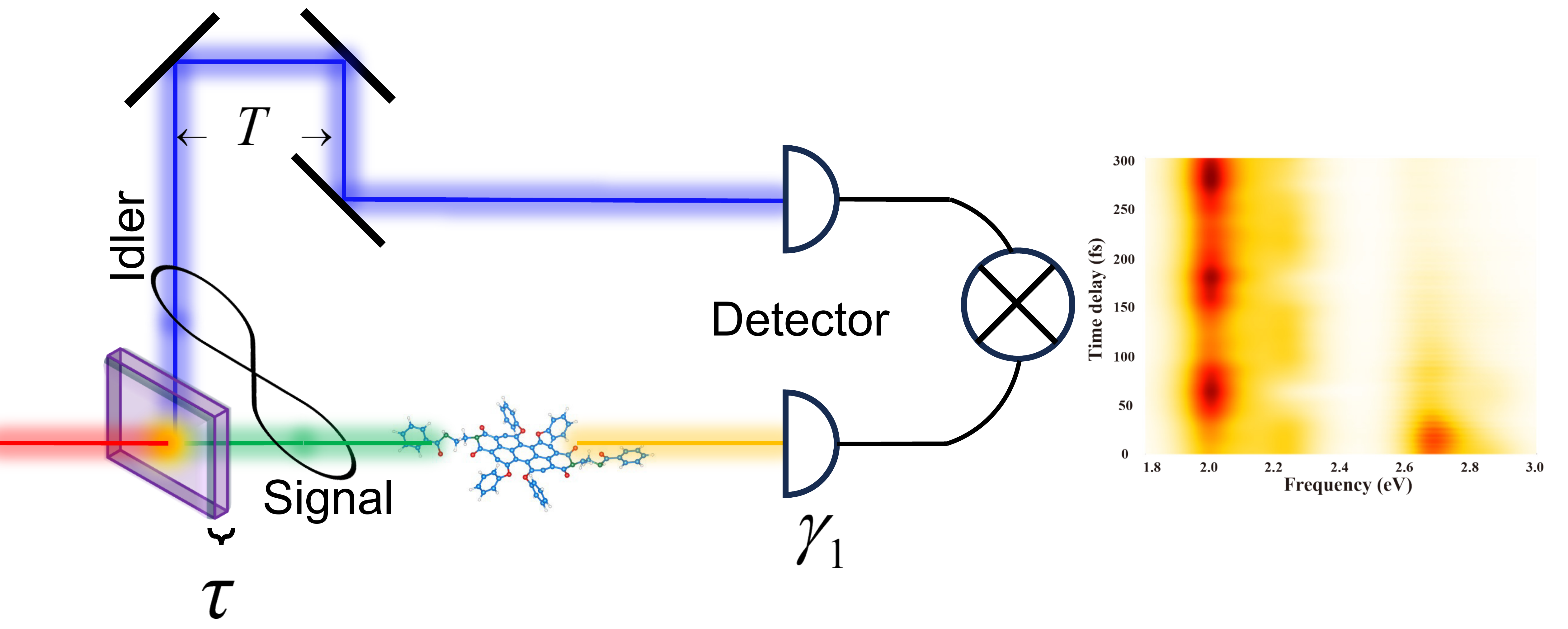}

\end{minipage}%

\end{document}